\documentclass[11pt,preprintnumbers,nofootinbib,prd]{revtex4}
\usepackage{graphicx}
\usepackage{amsmath}
\usepackage{epsfig}
\usepackage{mathrsfs}
\newcommand{\be}{\begin{equation}}
\newcommand{\ee}{\end{equation}}
\newcommand{\ba}{\begin{eqnarray}}
\newcommand{\ea}{\end{eqnarray}}
\newcommand{\notE}{E\kern-0.6em\hbox{/}\kern0.05em}
\newcommand{\notEt}{E_{T}\kern-1.21em\hbox{/}\kern0.45em}

\def\bi{\begin{itemize}}
\def\ei{\end{itemize}}

\def\P{P_{\rm eff}}
\begin{document}

\begin{flushright}
MCTP-07-43\\
UCB-PTH-08/01\\
\end{flushright}

\title{The $G_2$-MSSM - \\ An $M$ Theory motivated model of Particle Physics}
\author{Bobby S. Acharya}
\affiliation{Abdus Salam International Centre for Theoretical
Physics, Strada Costiera 11, Trieste, Italy\\and\\INFN, Sezione di
Trieste, Italy}
\author{Konstantin Bobkov}
\author{Gordon L. Kane}
\author{Jing Shao}
\affiliation{Michigan Center for Theoretical Physics, University
of Michigan, Ann Arbor, MI 48109, USA}
\author{Piyush Kumar}
\affiliation{Department of Physics, University of California,
Berkeley, CA 94720 USA\\and\\Theoretical Physics Group, Lawrence
Berkeley National Laboratory, Berkeley, CA 94720 USA}

%{\large Gordon L. Kane$^1$\footnote[1]{Email: gkane@umich.edu},
% Piyush Kumar$^1$\footnote[2]{Email: kpiyush@umich.edu},
%Joseph D. Lykken$^2$\footnote[3]{Email: lykken@fnal.gov},
% Ting T. Wang$^1$\footnote[4]{Email: tingwang@umich.edu}} \\
%\vspace{1.cm} \renewcommand{\thefootnote}{\arabic{footnote}} {\it
%1. Michigan Center for Theoretical Physics \\ Ann
%  Arbor, MI 48109, USA \\
%2. Fermi National Accelerator Laboratory\\ P.O. Box 500, Batavia, IL 60510, USA}\\
\vspace{0.5cm}

%\date{\today}

\vspace{0.3cm}

\begin{abstract}
We continue our study of the low energy implications of
$M$ theory vacua on $G_2$ manifolds, undertaken in
\cite{Acharya:2007rc,Acharya:2006ia}, where it was shown that the
moduli can be stabilized and a TeV scale generated,
with the Planck scale as the only dimensionful
input. A well-motivated phenomenological model - the $G_2$-MSSM,
can be naturally defined within the above framework. In this
paper, we study some of the important phenomenological features of
the $G_2$-MSSM. In particular, the soft supersymmetry breaking
parameters and the superpartner spectrum are computed. The
$G_2$-MSSM generically gives rise to light gauginos and heavy
scalars with wino LSPs when one tunes the cosmological constant.
Electroweak symmetry breaking is present but fine-tuned. The
$G_2$-MSSM is also naturally consistent with precision gauge
coupling unification. The phenomenological consequences for
cosmology and collider physics of the $G_2$-MSSM will be reported
in more detail soon.
\end{abstract}

\maketitle
\newpage
\tableofcontents

\vspace{0.5cm}

\vspace{0.3cm}

\newpage \vspace{-1.2cm} \newpage

\section{Introduction and Summary}

This year we will enter the LHC era. The LHC experiments will likely dominate
particle physics for many years. String/$M$ theory phenomenology
has to address many challenges, the most important of which are related to
dynamical issues, such as supersymmetry breaking, moduli stabilization and
explaining the Hierarchy between the Electroweak and Planck scales, etc. Successfully
addressing these opens the possibility
to construct models of particle
physics beyond the Standard-Model (SM) within the framework of string/$M
$ theory and study them to the extent that testable predictions for real
observables at the LHC (as well as other branches of particle physics and
cosmology) can be made. Different approaches will in general give rise to
different patterns of signatures; this could in principle distinguish among
different possibilities \cite{Kane:2006yi,Kane:2007pp}. In carrying out this
program in string/$M$ theory, it has to be kept in mind that the properties
of beyond-the-Standard-Model (BSM) particle physics models are intimately
connected to the dynamical issues mentioned above. This is because the
masses and couplings of the particle physics models depend on the properties
of the vacuum (or class of vacua) of the underlying string/$M$ theory
constructions, including the values of the moduli in the vacuum.

Substantial progress has been made in the past few years towards addressing
the dynamical issues of moduli stabilization, supersymmetry breaking and
explaining the Hierarchy, within various corners of the entire $M$ theory
moduli space in a reliable manner. The most popular one is the Type IIB
corner \cite{Kachru:2003aw}.
There have also been explicit semi-realistic
constructions of the visible sector within the Type IIB setup. While there
still do not exist explicit realizations incorporating mechanisms for
solving all the above mentioned dynamical problems as well as incorporating
a realistic visible sector within one particular construction, it is
possible to consider {\it frameworks} in which the relevant
effects of the underlying mechanisms may reasonably be assumed to exist in a
self-consistent manner. For example, within Type IIB string theory, one
popular {\it framework} is  warped flux
compactification of Type IIB string theory to four dimensions in the
presence of $D3$, anti ${D3}$ and $D7$-branes. The closed string fluxes
stabilize the dilaton and complex structure moduli whilst non-perturbative
effects stabilize the K\"{a}hler moduli. The anti-${D3}$-branes reside at the
end of the warped throat and break supersymmetry while the visible sector
(composed of $D3/D7$-branes) resides in the bulk of the internal manifold.
Supersymmetry breaking at the end of the throat is mediated to the visible
sector by the (higher dimensional) gravity multiplet, i.e. the moduli. In
this {\it framework}, also known as \textquotedblleft mirage
mediation", assuming the existence of all the relevant effects of the
underlying microscopic string compactification, one can incorporate them in
a self-consistent manner and then study the phenomenological consequences
for low energy observables. Such studies have been carried out in the
literature \cite{Choi:2005ge}.

The object of interest in this paper is a {\it framework} arising
from another weakly coupled limit: the low energy limit of $M$
theory. It was shown in \cite{Acharya:2007rc,Acharya:2006ia} that,
with reasonable assumptions about microscopic structure of the
underlying
construction, $\mathcal{N}$=1 fluxless compactifications of $M$ theory on $%
G_2$ manifolds can generate the hierarchy between the electroweak
and Planck scales and stabilize all the moduli in a dS vacuum.
This framework offers the possibility for studying the
consequences for low energy phenomenology. That is the goal of
this paper. In a series of companion papers, more detailed
applications for cosmology and collider physics will be presented.

The results obtained from the analysis of vacua within the above
framework are quite interesting and lead us to define a class of
particle physics  models which we call the $G_2$-MSSM. Many detailed properties
of these models are derived from $M$ theory, though we also add some necessary
assumptions due to theoretical uncertainties about the $M$ theory framework.

The $G_2$-MSSM models have a distinctive
spectrum.
One finds that, at the
compactification scale ($\sim M_{\mathrm{unif}}$), the gauginos
are light ($\lesssim 1$ TeV) and are suppressed compared to the
trilinears, scalar and  Higgsino masses which are roughly equal to
the gravitino mass ($\sim 30-100$ TeV).
At the electroweak scale,
the lightest top squark turns out to be significantly lighter than
the other squarks ($\sim 1-10$ TeV) because of RGE running. In
addition, there are significant finite threshold corrections to
bino and wino masses from the large Higgsino mass. Radiative
electroweak symmetry breaking is generic and $\tan \beta $ is
naturally predicted from the structure of the high scale theory to
be of $\mathcal{O}$(1). \ (Theoretical
predictions of $\tan \beta $ are fairly rare). The value of $m_{Z}$ is
fine-tuned, however, implying the existence of the
Little-hierarchy problem, which, because of the larger scalar masses
is worse than the usual little hierarchy.

The class of vacua within this framework also tend to be
consistent with precision gauge unification
\cite{Friedmann:2002ty}. The LSP usually turns out to be wino for
solutions consistent with precision gauge unification. Though the
thermal relic abundance of wino LSPs is quite low, an analysis of
the cosmological evolution of the moduli shows that non-thermal
production of wino LSPs provides about the correct amount of
dark-matter\footnote{Preliminary.}.
The collider phenomenology of the vacua within this
framework is also quite distinctive, giving rise to many $b$-jets
from multi-top production and short charged track stubs from the
decay of the wino-like chargino.

The paper is organized as follows. In the rest of this section, the results
obtained in \cite{Acharya:2006ia,Acharya:2007rc} are summarized and the
assumptions about the microscopic structure needed to define the framework
and make contact with low energy physics are specified. In particular, $G_2$%
-MSSM vacua, whose detailed phenomenology will be studied, are motivated and
introduced. Readers interested only in phenomenological results may skip
section I. In section \ref{soft-unif}, the computation of soft supersymmetry
breaking parameters at the unification scale consistent with all relevant
constraints is presented. Section \ref{spectra} deals with RG evolution,
calculation of the superpartner spectrum and Electroweak Symmetry Breaking
(EWSB). Section \ref{gaugino-gcu} studies precision gauge coupling
unification and its relation to gaugino masses. In section \ref{LSP}, we
analyze the nature of the LSP and its connection to the Dark Matter (DM)
relic density and the cosmological moduli problem. A brief discussion about
benchmark spectra and collider phenomenology of the class of vacua within
this framework appears in section \ref{features}. We conclude in section \ref%
{conclude}. This is followed by an Appendix discussing some
technical details about the computation of the quantity $\P$ that
enters in tuning the cosmological constant and in the gaugino mass
suppression, constraints on ``microscopic" parameters, and
threshold corrections to gaugino masses from heavy states.

\subsection{Summary of Results about $G_2$ Vacua}

\label{summary} In order to make our discussion of $G_2$
phenomenology self-contained, it is helpful to summarize the
essential results for the fluxless $M$ theory de Sitter vacua
described in \cite{Acharya:2006ia,Acharya:2007rc} and explain our
conventions and notation. Readers interested only in
phenomenological results may skip this section. $M$ theory
compactifications on singular $G_2$ manifolds are interesting in
the sense that they give rise to $\mathcal{N}=1$ supersymmetry in
four dimensions with non-Abelian gauge groups and chiral fermions.
The non-Abelian gauge fields are localized along three-dimensional
submanifolds of the seven extra dimensions whereas chiral fermions
are supported at points at which there is a conical singularity
\cite{Acharya:1998pm,Atiyah:2001qf,Acharya:2001gy,Acharya:2004qe}.
As explained in the introduction, in order to look at
phenomenological consequences of these compactifications in a
reliable manner, one has to address the dynamical issues of moduli
stabilization, supersymmetry breaking and generation of the
Hierarchy.

As explained in \cite{Acharya:2006ia,Acharya:2007rc}, one is interested in
the zero flux sector since then the moduli superpotential is entirely
non-perturbative. This is crucial for both stabilizing the moduli and
generating the Hierarchy naturally as we will review. Fluxes generate a large
superpotential and, unless there is a mechanism to obtain an exponentially large
volume of the extra dimensions, $G_2$ compactifications with flux will not
generate a small mass scale, such as the TeV scale.

We assume that the $G_{2}$ manifolds which we consider have
singularities giving rise to two non-Abelian, asymptotically free
gauge groups. This implies that they undergo strong gauge dynamics
at lower energies leading to the generation of a non-perturbative
superpotential. At least one of the hidden sectors is assumed to
contain light charged matter fields $\mathcal{Q}$ and
$\tilde{\mathcal{Q}}$ (with $N_{f}<N_{c}$) as well. There could be
other matter fields which are much heavier and decouple well above
the corresponding strong coupling scale. Thus in the
minimal\footnote{
More complicated situations are possible, some of them are discussed in \cite%
{Acharya:2006ia,Acharya:2007rc}.} framework, one has two hidden sectors
living on three-manifolds with gauge group $G_{a}\times G_{b}$ undergoing
strong gauge dynamics, one of them having a pair of massless charged matter
fields transforming in the (anti)fundamental representation of the gauge
group. Of course, in addition, it is assumed that there is another
three-manifold on which the observable sector gauge theory with the appropriate
chiral matter content lives. This will be discussed more in the next
subsection.
This set of assumptions about the
the compactification manifold gives a working definition of the {\it framework}.

The $\mathcal{N}=1$ supergravity theory obtained in four
dimensions is then characterized by the following hidden sector
superpotential:
\begin{eqnarray}  \label{super}
W &=&m_p^3\left(C_1\,P\,\phi^{-(2/P)}\,e^{ib_1f_1}+C_2\,
Q\,e^{ib_2f_2}\right); \;b_1=\frac{2\pi}{P},\,b_2=\frac{2\pi}{Q}
\end{eqnarray}
Here
$\phi\equiv\det(\mathcal{Q}\tilde{\mathcal{Q}})^{1/2}=(2\mathcal{Q}
\tilde {\mathcal{Q}})^{1/2}$ is the effective meson field (for one
pair of massless quarks) and $P$ and $Q$ are proportional to one
loop beta function coefficients of the two gauge groups which are
completely determined by the gauge group and matter
representations. For concreteness we can consider the gauge group
to be $SU(Q) \times SU(P+1)$ with one vector like family of quarks
charged under $SU(P+1)$. The normalization constants $C_1$ and
$C_2$ are calculable, given a particular $G_2$-manifold. $f_{1,2}$
are the (tree-level) gauge kinetic functions of the two hidden
sectors which in general are different from each other.
Schematically, the superpotential of each hidden sector is just
equal to the strong coupling scale of the the corresponding gauge
theory, i.e. $W \sim \Lambda_1^3 + \Lambda_2^3$. The vacuum
structure of the supergravity theory with this superpotential is
quite rich, but in general can only be studied numerically. A
special case exists however, when it is possible to study the
vacua semi-analytically. This is when the two three-manifolds on
which the hidden sector gauge fields are localised are in the same
homology class, which in terms of gauge kinetic function then
implies:
\begin{eqnarray}
f_1=f_2\equiv f_{\rm hid}=\sum_{i=1}^{N}\,N_i\,z_i; \;
z_i=t_i+is_i.
\end{eqnarray}
In the above equation, $s_i$ are the $N$ geometric moduli of the $G_2$ manifold (intuitively,
these characterise the
sizes of the 3-cycles in the $G_2$ manifold), while $t_i$ are the axionic
components coming from the 3-form field $C_{IJK}$ of eleven-dimensional
supergravity. The $N_i$ are integers which are determined by the homology class
of the hidden sector 3-cycles.

The supergravity potential is fully specified once the K\"ahler potential
is given.
The K\"{a}hler potential for matter fields in general is hard to
compute from first principles. However, owing to the fact that
matter fields are localized at points inside the seven dimensional
manifold $V_7$, it is reasonable to assume that the matter
K\"{a}hler potential is approximately canonical at leading order.
Then, the K\"{a}hler
potential is given by:
\begin{equation}\label{kahler}
K/m_{p}^{2}=-3\ln (4\pi ^{1/3}V_{7})+\bar{\phi}\phi
\end{equation}
where $V_{7}\equiv \frac{\mathrm{Vol}(X)}{l_{11}^7}$ is the volume of the $%
G_{2}$ manifold $X$ in units of the eleven-dimensional Planck length $l_{11}$%
, and is a homogenous function of the $s_{i}$ of degree $7/3$. A
simple and reasonable ansatz therefore is $V_{7}=\prod_{i=1}^{N}%
\,s_{i}^{a_{i}}$ with $a_{i}$ positive rational numbers subject to the
constraint $\sum_{i=1}^{N}a_{i}=\frac{7}{3}$ \cite{Acharya:2005ez}.
Many qualitative results about
moduli stabilization do not seem to rely on this special form of $V_{7}$, but
this form of $V_7$ is useful since it gives an $N-1$ parameter family of
K\"{a}hler potentials consistent with $G_2$-holonomy, which are tractable.
In a basis in which the K\"{a}hler potential is given by (\ref%
{kahler}), the gauge kinetic function is generically a function of
all the moduli, i.e. $N_{i}\neq 0,i=1,2..,N$.

In general
the scalar potential of the supergravity theory determined by $W$ and $K$
is a reasonably complicated function of all the moduli.
Therefore, one expects to find isolated meta-stable minima, which
indeed turns out to be the case, as explained in detail in \cite{Acharya:2007rc}.
The values of the moduli at the
minima are completely determined by the {\it microscopic constants}\footnote{
These are called "microscopic" because they determine the
effective lagrangian at the compactification ($\sim
M_{\mathrm{unif}}$) scale.}  -
$\{a_{i},N_{i},C_{1},C_{2},P,Q,N;\;i=1,2...N\}$ which characterize
the framework. Given a particular $G_2$-manifold consistent with our
assumptions, all of these
constants are calculable in principle. Therefore, given a {\it particular}
$G_2$-manifold within the framework, one obtains a {\it particular} set of microscopic
constants and a {\it particular} 4d ${\cal N}=1$ supergravity theory.

To find the minima of the moduli potential V explicitly, one first
stabilizes the axionic components of the complex moduli and the
phase of $\phi $. Then one minimizes the potential with respect to
$s_{i}$ and $|\phi | $, which leads to $N+1$ equations $\partial
_{s_{i}}V=0$ and $\partial _{|\phi |}V=0$ (for $N$ moduli). To
solve these equations analytically, we consider the class of
solutions in which the volume of the hidden sector three-manifold
$V_{\hat{Q}}$ supporting the hidden sector gauge groups is large.
This allows us to reduce the first set of $N$ equations into just
two simple equations, which can be solved order by order in a
$1/V_{\hat{Q}}$ expansion. Physically, this expansion can be
understood as an expansion in terms of the small gauge coupling of
the hidden sector - $(\alpha _{0})_{\mathrm{hid}}$ which is
self-consistent since our hidden sectors are assumed to be
asymptotically free. The solution corresponding to a metastable
minimum with spontaneously broken supersymmetry\ is given by
\begin{eqnarray}
s_{i} &=&\frac{a_{i}}{N_{i}}\frac{3}{14\pi }\frac{P_{\mathrm{eff}}\,Q}{Q-P}+%
\mathcal{O}(\P^{-1}),  \label{sol-1} \\
|\phi |^{2}
&=&1-\frac{2}{Q-P}+\sqrt{1-\frac{2}{Q-P}}+\mathcal{O}(\P^{-1}),
\label{sol-2}
\end{eqnarray}
where $P_{\mathrm{eff}}\equiv P\ln (C_{1}/C_{2})$. The natural
values of $P$ and $Q$ are expected to lie between ${\cal O}(1)$
and ${\cal O}(10)$. It is easy to see that a large
$P_{\mathrm{eff}}$ corresponds to small $\alpha $ for the hidden
sector
\begin{equation}
(\alpha _{0}^{-1})_{\mathrm{hid}}=\mathrm{Im}(f_{\mathrm{hid}})\approx \frac{%
Q}{2\pi (Q-P)}P_{\mathrm{eff}}
\end{equation}
implying that the expansion is effectively in
$P_{\mathrm{eff}}^{-1}$. The $\phi $ dependence of the potential
at the minimum is essentially
\begin{equation}
V_{0}\sim m_{3/2}^{2}M_{P}^{2}\left[ |\phi |^{4}+\left( \frac{4}{Q-P}+\frac{%
14}{P_{\mathrm{eff}}}-3\right) |\phi |^{2}+\left( \frac{2}{Q-P}+\frac{7}{P_{%
\mathrm{eff}}}\right) \right]
\end{equation}
Therefore, the vacuum energy vanishes if the discriminant of the above
expression vanishes, i.e. if
\begin{equation}\label{tune}
P_{\mathrm{eff}}=\frac{28(Q-P)}{3(Q-P)-8}.
\end{equation}
The above condition is
satisfied when the contribution from the $F$-term of the meson field
($F_{\phi }$) to the scalar potential cancels that from the
$-3m_{3/2}^{2}$ term. In this vacuum, the $F$-term of the moduli
$F_{i}$ are much smaller than $F_{\phi }$. In fact, all the $F_{i}$
vanish at the leading order in the $1/V_{\hat{Q}}$ expansion. For
$Q-P\leq 2$, there is no solution as either $s_{i}$ or $|\phi
|^{2}$ become negative. The first non-trivial solution occurs when
$Q-P=3$ for which $P_{\mathrm{eff}}=84$ is required to get a
vanishing vacuum energy (to leading order). The appearance of the
integer $84$ is closely related to the the dimensionality of
the $G_{2}$ manifold (which is 7) and that of the three-manifold
(which is 3). Other choices of $Q-P$ are also possible
theoretically, but are not interesting because of the following
reasons: a) the corresponding solutions, if they are to remain in
the supergravity regime, require the $G_{2}$ manifold to have a
rather small number of moduli $N$, since $N <
\frac{14Q}{(3(Q-P)-8)\pi}$\cite{Acharya:2007rc}. It is unlikely that
$G_2$-manifolds with such few moduli are capable of containing the
$MSSM$ spectrum, which has more than a hundred relevant couplings.
b) these
solutions generically lead
to an extremely high susy breaking scale as will be seen in section \ref%
{gravitino} on the gravitino mass. So phenomenologically interesting $G_{2}$
compactifications arise only for the case $Q-P=3$ and $P_{\mathrm{eff}}=84$.

Some comments on the requirement of $P_{\mathrm{eff}}=84$ are in order.
First of all, it is only a leading order result for the potential in $%
(\alpha _{0}^{-1})_{\mathrm{hid}}$ expansion. In fact, including
higher order $(\alpha _{0}^{-1})_{\mathrm{hid}}$ corrections leads
to the requirement $P_{\mathrm{eff}}\approx 83$
\cite{Acharya:2007rc}. The potential will also receive higher
order corrections in the $M$ theory expansion which will change
the requirement for $P_{\mathrm{eff}}$ (probably by a small
amount). One important good feature of the framework is that these
higher order corrections to the vacuum energy have little effect
on phenomenologically relevant quantities. Therefore, it is
sufficient to tune the vacuum energy to leading order as long as
one is interested in phenomenological consequences. From a
microscopic point of view, however, there are two issues - a) Is
it possible to realize a large value of $P_{\mathrm{eff}}$ from
explicit constructions? and b) Can the values of
$P_{\mathrm{eff}}$ scan finely enough such that one can obtain the
observed tiny
value of the cosmological constant? Regarding a), one notices that $P_{%
\mathrm{eff}}$ depends on the detailed structure of the hidden sector and is
completely model dependent. For particular  realizations of the hidden
sector, $P_{\mathrm{eff}}$ can be computed. A detailed discussion about $P_{%
\mathrm{eff}}$ is given in Appendix \ref{Peff}. However computing $P_{%
\mathrm{eff}}$ in more general cases is difficult  because of our
limited knowledge of possible three-dimensional submanifolds of
$G_{2}$ manifolds. In our analysis, we have assumed that
three-manifolds exist for which it is possible to obtain a large
$P_{\mathrm{eff}}$. The situation regarding b) is even less known.
This is because very little is known in general about the set of
all compact $G_{2}$ manifolds. Duality arguments do suggest that the
set of compact $G_2$-manifolds is larger than the space of compact
Calabi-Yau threefolds. Unfortunately, there do not curently exist any
concrete ideas about that space either!
In our work, we have assumed effectively that the
space of $G_{2} $ manifolds scans $P_{\mathrm{eff}}$ finely enough
such that vacua exist with values of the cosmological constant as
observed.

\subsection{The Observable Sector - Introducing the $G_2$-MSSM}

\label{G2MSSM}

In these compactifications, as mentioned earlier, the
observable sector gauge theory resides on a three-manifold different from the
one supporting the hidden sector. The observable sector three-manifold
is assumed to contain conical singularities at which chiral matter is
supported. Since two three-manifolds in a seven dimensional manifold
generically do not intersect each other, this implies that the supersymmetry
breaking generated by strong gauge dynamics in the hidden sector is generically
mediated to the visible sector by the (higher dimensional) gravity
multiplet. This gives rise to gravity (moduli) mediation. However, as will
be seen later, anomaly mediated contributions  will also play an important role for
the gaugino masses.

In our analysis henceforth, we will assume a GUT gauge group in
the visible sector which is broken to the SM gauge group, with at least an
MSSM chiral spectrum, by background gauge fields (Wilson lines). This assumption is
well motivated by considering the duality to the $E_8 \times E_8$ heterotic
string on a Calabi-Yau threefold.
For simplicity, we will present our results for the $SU(5)$ GUT group breaking
to the SM group and just an MSSM chiral spectrum, but all our results should hold
for other GUT groups breaking in the same way as well.

%*********************
%*************

To summarize, the full low energy $\mathcal{N}=1$ Supergravity theory of the
visible and hidden sectors at the compactification scale ($\sim M_{\mathrm{%
unif}}$) is defined by the following:%
\begin{eqnarray}  \label{eq:KW}
K/m_p^2 &=& \left(-3\ln(4\pi^{1/3}V_7) + \bar{\phi}\phi\right) + \tilde{K}_{%
\bar{\alpha}\beta}(s_i) \,\bar{\Phi}_{MSSM}^{\bar{\alpha}}
\Phi^{\beta}_{MSSM} + (Z(s_i)\,H_uH_d+h.c.)+ ...  \notag \\
W &=& m_p^3\left(C_1\,P\,\phi^{-(2/P)}\,e^{ib_1f_1}+C_2\,
Q\,e^{ib_2f_2}\right) + Y^{\prime }_{\alpha\beta\gamma}\, {\Phi}%
^{\alpha}_{MSSM}{\Phi}^{\beta}_{MSSM}{\Phi}^{\gamma}_{MSSM}  \notag \\
f_1 &=& f_2 \equiv f_{\rm hid}= \sum_i N^iz_i;\;\;\;\;
{\rm Im}(f^0_{vis})= \sum_i N^i_{vis}\,s_i\equiv V_{%
\hat{Q}_{vis}}
\end{eqnarray}
The visible sector is thus characterized by the K\"{a}hler metric $\tilde{K}%
_{\bar{\alpha}\beta}$ and un-normalized Yukawa couplings $Y^{\prime
}_{\alpha\beta\gamma}$ of the visible sector chiral matter fields ${\Phi}%
^{\alpha}_{MSSM}$ and the (tree-level) gauge kinetic function
$f^0_{vis}$ of the visible sector gauge fields. In addition, as is
generically expected in gravity mediation, a non-zero coefficient
$Z$ of the Higgs bilinear is assumed. In general there can also be
a mass term (${\mu}^{\prime }$) in the superpotential $W$, but as
explained in \cite{Witten:2001bf}, natural discrete symmetries can
exist which forbid it, in order to solve the doublet-triplet
splitting problem. However, the Giudice-Masiero mechanism in general
generates effective $\mu$ and $B\mu$
parameters of $\mathcal{O}(m_{3/2})$.

The K\"{a}hler metric $\tilde{K}_{\bar{\alpha}\beta }$ will be
discussed in section \ref{scalartri}. The visible sector gauge
couplings are determined by the gauge kinetic function $f_{vis}$
which is an integer linear combination of the moduli with the
integers determined by the homology class of the three-manifold
$\hat{Q}$ which supports the visible gauge group. Because of a
GUT-like spectrum, the MSSM gauge couplings are unified at
$M_{\mathrm{unif}}$ giving rise to the same $f^0_{vis}$. Since we
are assuming an MSSM visible sector below the unification scale,
one has to subject $ N_{vis}^{i}$ to the constraint that
$f^0_{vis}(M_{\mathrm{unif}})\equiv \alpha
_{unif}^{-1}(M_{\mathrm{unif}})\sim \mathcal{O}(25)$. The
Yukawa couplings in these vacua arise from
membrane instantons which connect singularities where chiral
superfields are supported (if some singularities coincide, there
could also be $\mathcal{O}(1)$ contributions). They are given by:
\begin{equation}
Y_{\alpha \beta \gamma }^{\prime }=C_{\alpha \beta \gamma }\,e^{i2\pi
\sum_{i}l_{i}^{\alpha \beta \gamma }z^{i}}  \label{Yukawas1}
\end{equation}
\noindent where $C_{\alpha \beta \gamma }$ is an ${\mathcal{O}}(1)$ constant
and $l_{i}^{\alpha \beta \gamma }$ are integers. Factoring out the phases,
the magnitude of the Yukawas can be schematically written as:
\begin{equation}
|Y|\sim |C|e^{-2\pi \,\vec{l}\cdot \vec{s}}
\end{equation}
The normalized Yukawas differ from the above by factors corresponding to
field redefinitions. Because of the exponential dependence on the moduli, it
is natural to obtain a hierarchical structure of Yukawa couplings as is
observed in nature. However, in general it is very difficult technically
to compute the Yukawa couplings quantitatively. Therefore,
for our phenomenological analysis, we will assume that the (normalized)
Yukawa couplings are the same as those of the Standard Model. This is
reasonable as in this work we are primarily interested in studying the
effects of supersymmetry breaking and electroweak symmetry breaking.

Since the moduli have been stabilized, the $F$-terms of the moduli
($F_i$) and the meson fields ($F_{\phi}$), which are the source of
supersymmetry breaking, can be computed explicitly in terms of the
microscopic constants. The expressions for $F_i$ and $F_\phi$ in
terms of these microscopic constants have been given explicitly in
\cite{Acharya:2007rc}. Since these $F$-terms and the quantities in
(\ref{eq:KW}) together determine the soft supersymmetry breaking
parameters, it becomes possible to express all the soft parameters
- gaugino masses, scalar masses, trilinears, $\mu$ and $B\mu$, in
terms of the microscopic constants. Thus, given a particular
$G_2$-manifold one obtains a particular set of microscopic
constants and thus a particular point in the MSSM parameter space.
The set of microscopic constants consistent with the framework of
$G_2$ compactifications and our assumptions thus defines a subset
of the MSSM, which we call the $G_2$-MSSM. How this works in
practice should become clear in the following sections. Formulae
(\ref{grav}), (\ref{scalarmass}), (\ref{trilinear}), (\ref{gauginohigh}),
(\ref{mu}) give the soft-breaking parameters at the unification
scale, in terms of the microscopic constants.

Before moving on to discussing the phenomenology of the $G_2$-MSSM vacua in
detail, it is worth noting that realistic $M$ theory vacua with a visible
sector larger than the MSSM, will give rise to additional,
different predictions for low energy
phenomenology in general, and LHC signatures in particular. Therefore, the
pattern of LHC signatures may help in distinguishing them. We hope to study
this issue in the future.

\section{$G_2$-MSSM soft supersymmetry breaking parameters at $M_{\mathrm{%
unif}}$}

\label{soft-unif}

Phenomenologically relevant physics quantities are not sensitive to all
of the microscopic constants. Instead, only certain parameters such as $Q$, $P$, $%
C_2$ and $\delta$ \footnote{This is defined in section
\ref{tree}.} as well as certain combinations of them, such as
$V_7$, $V_{\hat{Q}_{vis}}$ and $P_{\mathrm{eff}}$ are responsible
for relevant physics quantities. These combinations have to
satisfy various
constraints which make the procedure of moduli stabilization in \cite%
{Acharya:2007rc} valid and consistent. These conditions
give phenomenologically interesting consequences.
Four of the most important constraints are:

\begin{itemize}
\item The validity of the supergravity approximation requires $V_7 >  1$. We
call this the ``weak supergravity constraint''. In  its ``strong" form, one
could require all geometric moduli  $s_i$ to be greater than unity. The
supergravity constraint is  required because our solutions can be trusted
only in this regime.  There could be different solutions in other regimes
beyond the supergravity apporximation, but little is known about them.

\item Since current observations indicate a strong evidence for a dS vacuum
with a tiny cosmological  constant, the vacua are required to have positive
cosmological constant. In addition, the  microscopic constants are required to be such
that the cosmological constant can be tuned to be very small. We call this
the ``dS vacuum constraint".

\item Since it is known that, for an MSSM visible sector $\alpha_{\mathrm{unif}%
}^{-1} \sim \mathcal{O}(25)$, this implies  $V_{\hat{Q}_{vis}} \sim \alpha_{%
\mathrm{unif}}^{-1} \sim \mathcal{O}(25)$. We call this the ``unified
coupling constraint''.

\item $M_{11} > M_{\mathrm{unif}} \sim \mathcal{O}(10^{16}$ GeV), which is
required to make intrinsic $M$ theory  corrections to gauge couplings and
other parameters negligible  at and below $M_{\mathrm{unif}}$. We call this
the ``unification scale constraint''.
\end{itemize}

A discussion of parameters compatible with the ``supergravity constraint",
``dS vacuum constraint" and the ``unified coupling constraint" is given
in Appendix \ref{constraints}. In section \ref{gaugino-gcu}, it
will be shown that the ``unification scale" constraint can be naturally
satisfied within this framework. We will now proceed to discussing the
gravitino mass and the soft supersymmetry breaking parameters assuming that
sets of parameters can be found such that all the above constraints are
satisfied.

\subsection{Gravitino Mass}

\label{gravitino}

The bare gravitino mass in $\mathcal{N}=1$ supergravity can be computed as follows:
\begin{eqnarray}  \label{grav1}
m_{3/2}&\equiv&m_p^{-2}\,e^{K \over 2m_p^2}\,|W|
\end{eqnarray}
This quantity plays an important role in gravity mediated models of supersymmetry breaking
and sets the typical mass scale for couplings in the supergravity Lagrangian.
It is therefore useful to compute this quantity in detail in the $G_2$-MSSM.

As explained earlier, $|W|$ is generated by strong gauge dynamics
in the two hidden sectors, $W_{1,2}\sim (\Lambda_{1,2})^3$.
This implies that the gravitino mass can be
schematically expressed as:
\begin{eqnarray}
m_{3/2}\sim \frac{\Lambda^3}{m_p^2}
\end{eqnarray}
up to some factors of volume coming from the K\"{a}hler potential in (\ref%
{grav1}). More precisely, the gravitino mass is:
\begin{eqnarray}
m_{3/2}&=&m_p\frac{e^{\phi_0^2/2}}{8\sqrt{\pi}V_7^{3/2}}C_2 \Big|P\phi_0^{-%
\frac{2}{P}}-Q\Big|e^{-\frac{P_{\mathrm{eff}}}{Q-P}}  \notag \\
&\approx&m_p\frac{e^{\phi_0^2/2}}{8\sqrt{\pi}V_7^{3/2}}C_2 |Q-P|e^{-\frac{P_{%
\mathrm{eff}}}{Q-P}}  \label{grav}
\end{eqnarray}
where in the last line, $\phi_0 \approx 1$ is used. The exponential part in
the equation is roughly $\Lambda_{cond}^3$ in units of $m_p$. As seen from
above, the gravitino mass is effectively determined by four parameters: $%
\{P_{\mathrm{eff}}$, $Q-P$, $V_7$ and $C_2\}$. For $Q-P=3$, the gravitino
mass can be well approximated by
\begin{eqnarray}  \label{grav-approx}
m_{3/2}=(708~\mathrm{TeV})\times \left(C_2 V_7^{-3/2}e^{-\frac{P_{\mathrm{eff%
}}-83}{3}}\right)
\end{eqnarray}
For the case of zero cosmological constant $P_{\mathrm{eff}}=83$ \footnote{%
The upper limit $P_{\mathrm{eff}}=84$ obtained in the zeroth order of the $%
1/V_{\hat{Q}_{vis}}$ approximation is modified to $P_{\mathrm{eff}}=83$
after taking higher order corrections into account.}, the exponential is
unity and the gravitino mass is bounded from above by $708$ TeV. As will be
seen later, $V_7$ turns out to be typically in the range 10-100, implying
that $m_{3/2}$ naturally lies between 10 and 100 TeV. If one allows a dS
minimum with a large cosmological constant, $P_{\mathrm{eff}}$ can be
smaller than $84$ and the gravitino mass can become larger. For larger
values of $Q-P$, the $P_{\mathrm{eff}}$ required to tune the cosmological
constant (see eqn. (\ref{tune})) is smaller. For example, for $Q-P=4$, $P_{%
\mathrm{eff}}=28$ is required. Since the gravitino mass is exponentially
sensitive to $P_{\mathrm{eff}}$ (as seen from (\ref{grav-approx})), the
gravitino mass for $Q-P > 3$ turns out to be much larger then the TeV scale.
This is the reason for mainly considering the case $Q-P=3$.

\subsection{Scalars and Trilinears at $M_{\mathrm{unif}}$}

\label{scalartri}

The general expressions for the \emph{un-normalized} scalar masses and
trilinear parameters are given by \cite{Brignole:1997dp}: \label{m'A'}
\begin{eqnarray}
m^{\prime 2}_{\bar{\alpha}\beta} &=& (m_{3/2}^2 + V_0)\,\tilde{K}_{\bar{%
\alpha}\beta} - e^{\hat K}F^{\bar{m}}(\partial_{\bar{m}}\partial_n\,\tilde{K}%
_{\bar{\alpha}\beta}-\partial_{\bar{m}}\, \tilde{K}_{\bar{\alpha}\gamma}\,%
\tilde{K}^{\gamma\bar{\delta}}\partial_n\,\tilde{K}_{\bar{\delta}\beta})F^n
\\
A^{\prime }_{\alpha\beta\gamma} &=& \frac{\hat{W}^{\star}}{|\hat{W}|} e^{%
\hat{K}}\,F^m\,[\hat{K}_m\,Y^{\prime }_{\alpha\beta\gamma}+\partial_m
Y^{\prime }_{\alpha\beta\gamma}-(\tilde{K}^{\delta\bar{\rho}}\,\partial_n%
\tilde{K}_{\bar{\rho}\alpha}\,Y^{\prime }_{\alpha\beta\gamma}+ \alpha
\leftrightarrow \gamma + \alpha \leftrightarrow \beta)]  \notag
\end{eqnarray}
In order to determine physical implications, however, one has to canonically
normalize the visible matter K\"{a}hler potential $K_{visible} = \tilde{K}_{%
\bar{\alpha}\beta}\Phi^{\bar{\alpha}}\Phi^{\beta}+...$, which is achieved by
introducing a normalization matrix $\mathcal{U}$ :
\begin{eqnarray}  \label{norm}
\Phi \rightarrow \mathcal{U} \cdot \Phi, \;\;\; s.t. \;\;\; \mathcal{U}%
^{\dag}\tilde{K}\mathcal{U} = 1.
\end{eqnarray}
The $\mathcal{U}s$ are themselves only defined up to a unitary
transformation, i.e. $\mathcal{U}^{\prime }=\mathcal{U}\cdot\mathcal{N}$ is
also an allowed normalization matrix if $\mathcal{N}$ is unitary. The \emph{%
normalized} scalar masses and trilinears can then be written formally as:
\begin{eqnarray}
m^2_{\bar{\alpha}\beta} &=& (\mathcal{U}^{\dag}\cdot m^{\prime 2 }\cdot
\mathcal{U})_{\bar{\alpha}\beta} \\
\tilde{A}_{\alpha\beta\gamma} &=& \mathcal{U}_{{\alpha}\alpha^{\prime }}%
\mathcal{U}_{{\beta}\beta^{\prime }}\mathcal{U}_{{\gamma}\gamma^{\prime }}
A^{\prime }_{{\alpha}^{\prime }{\beta}^{\prime }{\gamma}^{\prime }}  \notag
\end{eqnarray}

More precisely, the scalar masses can be written as:
\begin{eqnarray}  \label{scalars1}
m^2_{\bar{\alpha}\beta} &=& (m_{3/2}^2 + V_0)\,\delta_{\bar{\alpha}\beta} -
\mathcal{U}^{\dag}\Gamma_{\bar{\alpha}\beta} \mathcal{U} \\
\Gamma_{\bar{\alpha}\beta} &\equiv& e^{\hat K}F^{\bar{m}}(\partial_{\bar{m}%
}\partial_n\,\tilde{K}_{\bar{\alpha}\beta}-\partial_{\bar{m}}\, \tilde{K}_{%
\bar{\alpha}\gamma}\,\tilde{K}^{\gamma\bar{\delta}}\partial_n\,\tilde{K}_{%
\bar{\delta}\beta})F^n  \notag
\end{eqnarray}
Thus, when the cosmological constant has been tuned to be small, the scalar
masses generically have a flavor universal and flavor diagonal contribution
equal to $m_{3/2}^2$ from the first term in (\ref{scalars1}) and a flavor
non-universal and flavor non-diagonal contribution from the second term in (%
\ref{scalars1}). In order to estimate the size of the
non-universal and non-diagonal contributions, one has to know
about the moduli and meson dependence of the visible sector
K\"{a}hler metric. This dependence of the matter K\"{a}hler metric
is notoriously difficult to compute in generic string and $M$
theory vacua, and the vacua under study here are no exception.
Therefore, it is only possible to proceed by making reasonable
assumptions.
Under our assumptions about the meson field kinetic term,
the only contribution to
the non-universal and non-diagonal terms in (\ref{scalars1}) comes from the $%
F$ terms of the moduli $F_i$. Since $F_i \ll F_{\phi}$,
the non-universal contributions are
negligible compared to the universal and diagonal contribution. Thus,
\begin{eqnarray} \label{scalarmass}
m^2_{\bar{\alpha}\beta} \approx m_{3/2}^2\,\delta_{\bar{\alpha}\beta}
\end{eqnarray}
This implies that flavor changing neutral currents (FCNCs) are
adequately suppressed. The fact that the scalar masses are
are roughly equal to the gravitino mass can be traced to the
non-sequestered nature of the K\"{a}hler potential in
(\ref{eq:KW}). In the absence of fluxes, the $G_2$
compactifications considered here do not have any warping, which
implies that one generically does not have sequestering in these
compactifications \cite{Kachru:2007xp}. Since the scalars are
heavy and also flavor universal at leading order, we expect that
no significant signals should occur for observables from loops
with sleptons or squarks, in particular for rare flavor violating
decays such as $\mu \rightarrow e \gamma$, $K \rightarrow \pi \nu
\bar{\nu}$, $b \rightarrow s \gamma$, etc, and also no significant
signal for $g_{\mu}-2$.

The computation of the trilinears also simplifies under the above
assumptions. Again, since the un-normalized Yukawa couplings and
the visible sector K\"{a}hler metric are not expected to depend on
the meson field, the dominant contribution to the trilinears comes
from the first term in the expression for trilinears in
(\ref{m'A'}). Thus, one has:
\begin{eqnarray}
A^{\prime }_{\alpha\beta\gamma} \approx \frac{\hat{W}^{\star}}{|\hat{W}|} e^{%
\hat{K}}\,F^{\phi}\,\hat{K}_{\phi}\,Y^{\prime }_{\alpha\beta\gamma}
\end{eqnarray}
This implies that the normalized trilinear parameters are:
\begin{eqnarray}  \label{trilinear}
\tilde{A}_{\alpha\beta\gamma} &\approx& (\mathcal{U}_{{\alpha}\alpha^{\prime
}}\mathcal{U}_{{\beta}\beta^{\prime }}\mathcal{U}_{{\gamma}\gamma^{\prime
}})\, \frac{\hat{W}^{\star}}{|\hat{W}|}e^{\hat{K}}\,F^{\phi}\,\hat{K}%
_{\phi}\,Y^{\prime }_{\alpha^{\prime }\beta^{\prime }\gamma^{\prime }}
\notag \\
&\approx& e^{\hat{K}/2}\,F^{\phi}\,\hat{K}_{\phi}\,Y_{\alpha\beta\gamma}
\notag \\
&\approx& e^{-i\gamma_W}\,1.48\,m_{3/2}\,Y_{\alpha\beta\gamma}
\end{eqnarray}
Here we have used the fact that the normalized Yukawa couplings are given by
$Y_{\alpha\beta\gamma}=\frac{\hat{W}^{\star}}{|\hat{W}|}e^{\hat{K/2}} (%
\mathcal{U}_{{\alpha}\alpha^{\prime }}\mathcal{U}_{{\beta}\beta^{\prime }}%
\mathcal{U}_{{\gamma}\gamma^{\prime }}) \,Y^{\prime }_{\alpha^{\prime
}\beta^{\prime }\gamma^{\prime }}$ and that $e^{\hat{K/2}}\,F^{\phi}\,\hat{K}%
_{\phi} \approx  e^{-i\gamma_W}\,\sqrt{3}\,\phi_0\,m_{3/2} \approx
e^{-i\gamma_W}\,1.48\,m_{3/2}$\cite{Acharya:2007rc}. The reduced normalized
trilinear parameters have a particularly simple form:
\begin{eqnarray}
{A}_{\alpha\beta\gamma} \equiv \frac{\tilde{A}_{\alpha\beta\gamma}}{%
Y_{\alpha\beta\gamma}}\approx e^{-i\gamma_W}\,1.48\,m_{3/2}
\end{eqnarray}
Thus, we see that in $G_2$-MSSM vacua, the scalar masses and trilinears
are generically of $\mathcal{O}(m_{3/2})$.

\subsection{Gaugino Masses at $M_{\mathrm{unif}}$}

\label{gauginounif}

We now turn to gaugino masses. The computation of gaugino masses depends
less on our knowledge of the matter K\"{a}hler potential, therefore it is
possible to obtain quite detailed formulae.
In $G_2$ vacua the
tree-level gaugino masses are suppressed relative to the gravitino mass
unlike the scalars and trilinears. Therefore, other contributions such as
those from anomaly mediation and those from threshold effects arising from
integrating out heavy fields can be important. Schematically, one can write
\begin{eqnarray}  \label{gauginoschematic}
M_a (\mu) = M_a^{\mathrm{tree}}(\mu)+M_a^{\mathrm{AMSB}}(\mu)+M_a^{\mathrm{%
thres}}(\mu)
\end{eqnarray}
In the following we wish to compute each contribution at the unification
scale $M_{\mathrm{unif}}$. We study the case when the low
energy spectrum is that of the MSSM. As mentioned in section \ref{G2MSSM},
for concreteness we will assume an $SU(5)$ GUT group broken to the MSSM by a
discrete choice of Wilson lines for concreteness. This gives rise to a pair
of Higgs triplets whose effects should be properly taken into account. For
the case of a different GUT group breaking to the MSSM by Wilson lines,
there would be similar heavy particles whose effects should be taken into
account. As we will see, the results obtained will be the same for all GUT
groups as long as the low energy spectrum is that of the MSSM.

\subsubsection{Tree-level suppression of Gaugino masses}\label{tree}

The tree-level gaugino masses at the scale $\mu$ are given by
\cite{Brignole:1997dp}:
\begin{eqnarray}  \label{defn}
M_a^{\mathrm{tree}}(\mu)&=&\frac{g_a^2(\mu)}{8\pi}\,\left(\sum_{m,n}e^{\hat{K%
}/2}K^{m\bar{n}} F_{\bar{n}}\partial_m\,\mathrm{Im}f_a^0\right) \\
&=& \frac{g_a^2(\mu)}{8\pi}\,\sum_{i=1}^{N}e^{\hat{K}/2}K^{i\bar{i}} F_{\bar{%
i}}N_i^{vis}.
\end{eqnarray}
where $f_a^0$ is the tree-level gauge kinetic function of the
$a^{th}$ gauge group. As explained earlier, the tree-level gauge
kinetic function $f_{a}^0$ of the three gauge groups in the MSSM
are the same ($=f^0_{vis}$) because of the underlying GUT
structure. The tree-level gaugino mass at the unification scale
can then be computed in terms of microscopic parameters. The
details are provided in section VIIIA of \cite{Acharya:2007rc}.
Here, we write down the result:
\begin{eqnarray}  \label{gaugino33}
M_{a}^{\mathrm{tree}} (M_{\mathrm{unif}})&\approx&-\frac{e^{-i\gamma_W}\,\eta%
}{P_{\mathrm{eff}}}\Big(1+\frac{2}{\phi_0^2(Q-P)}+\mathcal{O}(P_{\mathrm{eff}%
}^{-1})\Big)\;m_{3/2}  \notag \\
\mathrm{where}\; \alpha^{-1}_{\mathrm{unif}} &=& \mathrm{Im}(f_{vis}^0) +
\delta;\;\;\eta = 1-\frac{\delta}{\alpha^{-1}_{\mathrm{unif}}}
\end{eqnarray}where $\gamma_W$ arises from the phase of the
$F$-term. The phases of the soft parameters will be discussed in
detail in section \ref{phases}. $\delta$ corresponds to threshold
corrections to the (unified) gauge coupling and will be discussed
more in section \ref{gaugino-gcu}. As seen from above, gaugino
masses are suppressed by $P_{\mathrm{eff}}$ relative to gravitino
mass. This property is independent of the details of the
K\"{a}hler potential for $\phi$ and the form of $V_7$.

Because the MSSM is obtained by the breaking of a GUT group by
Wilson lines, the gauge couplings of the MSSM gauge groups are
unified at the unification scale giving rise to a common
$\mathrm{Im} (f_{vis}^0)$. This implies that the tree level
gaugino masses at $M_{\mathrm{unif}}$ are also unified. In
particular, for $Q-P=3$ with a vanishing cosmological constant ($P_{\mathrm{%
eff}}=83$) and the K\"{a}hler potential given by (\ref{kahler}),
one has an explicit expression for the gaugino mass:
\begin{eqnarray}  \label{gauginotree}
M_{a}^{\mathrm{tree}}(M_{\mathrm{unif}}) &\approx& -
\frac{e^{-i\gamma_W}}{83
}\eta\left(1+\frac{2}{3\phi_0^2}+\frac{7}{83\phi_0^2}\right)\times
m_{3/2}
\notag \\
&\approx& \,e^{-i\gamma_W}\,\,\eta\, 0.024\,m_{3/2}
\end{eqnarray}
Here we have used the fact that for $Q-P=3$,
$\phi_0^2\approx0.73$. As explained in section \ref{summary}, a
large $P_{\mathrm{eff}}$ is required for the validity of our
solutions. Therefore, the parametric dependence on
$P_{\mathrm{eff}}^{-1}$ in Eq.(\ref{gaugino33}) indicates a large
suppression in gaugino masses. The precise numerical value of the
suppression may change if one considers a more general form of the matter K%
\"{a}hler potential since then the numerical factor multiplying $P_{\mathrm{%
eff}}^{-1}$ in (\ref{gaugino33}) may change in general. However as
long as the couplings for higher order terms in the matter
K\"{a}hler potential such as $(\bar\phi\phi)^2$ are sufficiently
small, a large numerical suppression is generic. In our analysis
henceforth, we will assume that to be the case.
%Equation (\ref{gaugino33}) is completely
%independent of the choice of integers $N_i^{sm}$ for the Standard
%Model gauge kinetic function as well as the integers $N_i$ for the
%hidden sector. It is also independent of the number of moduli $N$
%as well as the particular details of the internal manifold
%described by the rational numbers $a_i$ appearing in the
%K\"{a}hler potential (\ref{kahler}). These properties imply that
%relation (\ref{gaugino33}) is  universal for all $G_2$ holonomy
%compactifications consistent with our approximations, independent
%of many internal details of the manifold.

From a physical point of view, the suppression of gaugino masses
is directly related to the fact that the $F$-terms of moduli $F_i$
(in Planck units) are suppressed compared to $m_{3/2}$ and that
the gauge kinetic function $f_a^0$ in (\ref{defn}) only depends on
the moduli $s_i$. This implies that the $F$-term of the meson
field does not contribute in (\ref{gaugino33}). It is also useful
to compare the above result for the tree-level suppression of
gaugino masses in $G_2$ dS vacua with that of Type IIB dS vacua.
In KKLT and large volume type IIB compactifications, the moduli
$F$ terms also vanish in the leading order leading to a
suppression of tree-level gaugino masses, although for a different
reason - the flux contribution to the moduli $F$ terms cancels the
contribution coming from the non-perturbative superpotential \cite%
{Conlon:2006us}. Another difference is that the subleading contributions in
those Type IIB vacua are suppressed by the inverse power of the volume of
the compactification manifold. Note that a large associative three-cycle on
a $G_2$ manifold ($V_{\hat{Q}_{vis}}$) does \emph{not} translate into a
large volume compactification manifold. So, it is possible for $G_2$ vacua
to have a large $V_{\hat{Q}_{vis}}$ while still having a high
compactification scale.

\subsubsection{Anomaly contributions}

Since the tree-level gaugino mass is suppressed, the anomaly
mediated contributions become important and should be included. \
In our framework they are not suppressed. The general expression
for the anomaly contribution to gaugino masses at scale $\mu $ is
written as \cite{Bagger:1999rd}:
\begin{eqnarray}
M_{a}^{anom}(\mu ) &=&-\frac{g_{a}^{2}(\mu )}{16\pi ^{2}}\Big(%
-(3C_{a}-\sum_{i}C_{a}^{i})e^{\hat{K}/2}W^{\ast
}+(C_{a}-\sum_{i}C_{a}^{i})e^{{\hat K}/2}F^{m}\hat{K}_{m}  \notag  \label{anom1} \\
&+&2\sum_{i}(C_{a}^{i}e^{{\hat K}/2}F^{m}\partial _{m}\ln
(\tilde{K}_{i}))\Big),
\end{eqnarray}%
where $i$ runs over all the MSSM chiral fields. Again, since the K\"{a}hler
metric for visible sector matter fields is expected to be independent of the
meson field, the third term in (\ref{anom1}) is much smaller than the first
two. Thus, the anomaly mediated contribution can be simplified:
\begin{eqnarray}
M_{a}^{anom}(M_{\mathrm{unif}}) &\approx &-\frac{e^{-i\gamma _{W}}\,g_{a}^{2}%
}{16\pi ^{2}}\Big[b_{a}+\left( 1+\frac{2}{(Q-P)\,\phi _{0}^{2}}+\frac{7}{%
\phi _{0}^{2}\,P_{\mathrm{eff}}}\right) \,\left( -b_{a}^{\prime }(\phi
_{0}^{2}+\frac{7}{P_{\mathrm{eff}}})\right) \Big]\,m_{3/2}  \label{gauginoam}
\\
b_{a} &\equiv &-(3C_{a}-\sum_{i}C_{a}^{i});\;\;b_{a}^{\prime }\equiv
-(C_{a}-\sum_{i}C_{a}^{i})  \notag
\end{eqnarray}%
As seen from (\ref{gauginoam}), the anomaly mediated contributions
are not universal. Since the anomaly mediated contributions are
numerically comparable to the tree-level contributions, the
gaugino masses will be non-universal at the unification scale.
That the tree-level and anomaly mediated contributions are similar
in size seems to be accidental -- one is suppressed by $1/\P =
1/83$, the other by the loop factor $1/16\pi^2$, and these factors
are within a factor of two of each other.

\subsubsection{The complete Gaugino masses}

\label{complete}

In principle, there can also be threshold corrections to gaugino masses from
high scale physics and it is important to take them into account. In
general, a threshold correction to the gaugino masses at a scale $M_{th}$ is
induced by a threshold correction to gauge couplings by the following
expression \cite{Choi:2007ka}:
\begin{eqnarray}  \label{threshold}
\Delta M_a = g_a^2(M_{th})\,F^I\partial_I\,(\Delta\,f_a^{thresh})
\end{eqnarray}
In these $M$ theory compactifications, possible threshold corrections at
scales $\leq M_{\mathrm{unif}}$ can arise from the following:

\begin{itemize}
\item Generic heavy $M$ theory excitations $\Psi$ of $\mathcal{O}(M_{11})$ -
$\Delta f_a^{M\,theory}$.

\item 4D particles in the GUT multiplet with mass $\approx M_{\mathrm{unif}}$
- $\Delta f_a^{T,\tilde{T}}$.

\item Kaluza-Klein (KK) excitations of $\mathcal{O}(M_{\mathrm{unif}})$ - $%
\Delta f_a^{KK}$.
\end{itemize}

It turns out that threshold corrections to the gauge couplings from KK modes
are constants \cite{Friedmann:2002ty}. Therefore, from (\ref{threshold}),
they will \emph{not} give rise to any threshold correction to the gaugino
masses. As explained in appendix \ref{thresholdcorr}, corrections from
generic heavy $M$ theory states $T$ with mass $\sim M_{11}$ as well as from
4D heavy GUT particles with mass ($\sim M_{\mathrm{unif}}$) (such as Higgs
triplets in $SU(5)$), to the gaugino masses are also negligible. So, the
complete gaugino mass can be approximately written as:
\begin{eqnarray}  \label{gauginohigh}
M_{a}(M_{\mathrm{unif}})&\approx&-\frac{e^{-i\gamma_W}}{4\pi(\mathrm{Im}%
(f^0)+\delta)}\left\{b_a+\left(\frac{4\pi\,\mathrm{Im}(f^0)}{P_{\mathrm{eff}}%
}-b_a^{\prime }\phi_0^2\right)\Big(1+\frac{2}{\phi_0^2(Q-P)}+\frac{7}{%
\phi_0^2P_{\mathrm{eff}}}\Big)-\frac{7b_a^{\prime }}{P_{\mathrm{eff}}}
\right\}\, m_{3/2}  \notag \\
\mathrm{where}\quad \; b_1&=&33/5,\quad b_2=1.0, \quad b_3=-3.0, \quad
b_1^{\prime }=-\frac{33}{5}, \quad b_2^{\prime }=-5.0, \quad b_3^{\prime
}=-3.
\end{eqnarray}
The above analytical expression for gaugino masses is true up to
the first subleading order in the $1/{V}_{\hat{Q}}$ expansion. In
general, the full gaugino mass at the unification scale
(\ref{gauginohigh}) depends on the parameters
$\{\mathrm{Im}(f^{0})$, $Q-P$, $\delta$, $V_7$, $P_{\mathrm{eff}}$
and $C_2\}$. For the phenomenologically interesting case with
$Q-P=3$ and $P_{\mathrm{eff}}=83$, there are effectively only four
parameters: $\{\mathrm{Im}(f^{0}),\delta,V_7,C_2\}$. As seen from
(\ref{grav-approx}), $m_{3/2}$ is determined from the last two
parameters - $V_7$ and $C_2$. Therefore, the ratio of the gaugino
masses to the gravitino mass for the phenomenologically
interesting case of $Q-P=3, P_{\mathrm{eff}}=83$ just depends on
the parameters $\mathrm{Im}(f^{0})$ and $\delta$ which are subject
to the constraint
$\alpha^{-1}_{\mathrm{unif}}=\mathrm{Im}(f^{0})+\delta \approx
\mathcal{O}(25)$ (see section \ref{G2MSSM}).
\begin{figure}[h!]
\resizebox{15cm}{!}{\includegraphics{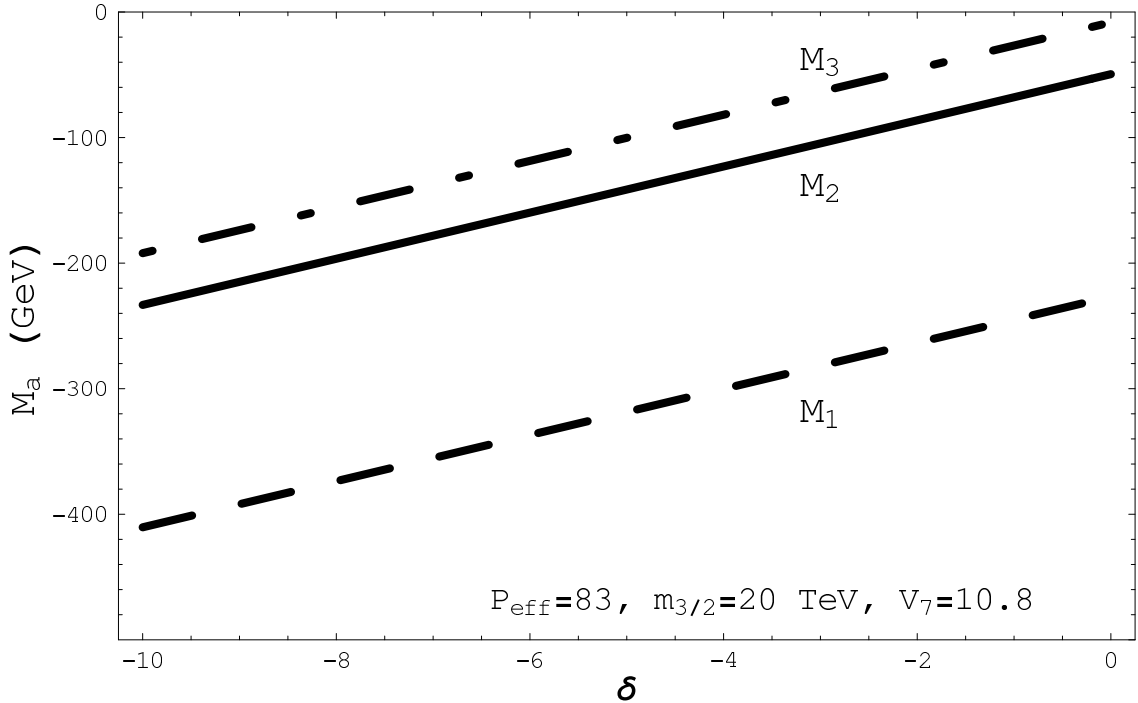}}
\caption{{\protect\footnotesize {The absolute value of the gaugino masses at
the unification scale are plotted as functions of threshold correction to
the gauge kinetic function $\protect\delta$ for $Q-P=3$, $P_{\mathrm{eff}}=83
$, $m_{3/2}$=20 TeV and $\protect\alpha_{unif}^{-1}=26.5$.}}}
\label{fig:gaugino-high1}
\end{figure}
The gaugino masses are plotted as a function of $\delta$ in Figure \ref%
{fig:gaugino-high1} for $m_{3/2}=20$ TeV. Notice that $M_2$ and $M_3$ are
smaller than $M_1$ by a factor of few at the unification scale. However, as
will be seen promptly, the gluino will still turn out to be the heaviest
gaugino because of RG effects.

\subsection{Phases, $\mu$ and $B\mu$.}

\label{phases}

Until now, we have not looked at the phases in detail. In general, phases
may have non-trivial phenomenological consequences. To count the number of
phases, it is helpful to go back to the expression for the superpotential:
\begin{eqnarray}
W=A_1 \phi^{\alpha} e^{ib_1 f}+A_2 e^{ib_2 f};\;\alpha\equiv
-\frac{2}{P}
\end{eqnarray}
where $A_1$ and $A_2$ and $\phi$ are complex variables with phases say ($%
\varphi_1$, $\varphi_2$, $\theta$). However the relative phase between the
first and second terms is fixed by the minimization of axions,
\begin{eqnarray}
\sin\Big((b_1-b_2){\vec N}\cdot{\vec
t}+\alpha\theta+\varphi_1-\varphi_2\Big)&=&0
\notag \\
\cos\Big((b_1-b_2){\vec N}\cdot{\vec
t}+\alpha\theta+\varphi_1-\varphi_2\Big)&=&-1
\end{eqnarray}
This can be seen in the Eq. (102) and (103) in \cite{Acharya:2007rc}. The
only phase left is $\gamma_W \equiv b_2 {\vec N}\cdot{\vec t}+\varphi_2$.

Let's start with the calculation of phases for the gauginos. From the
definition of the tree-level gaugino masses in (\ref{defn}), $M_a \propto {F}%
_{\bar m}\equiv\partial_{\bar m}{\bar W}+(\partial_{\bar m}K){\bar W}$.
Thus, the phase of the tree-level gaugino mass is $-\gamma_W$ (even though
the K\"{a}hler potential appears in the expression for gaugino masses, it is
real and hence does not contribute to the phase). There is an additional
contribution from anomaly-mediation which includes terms proportional to
either ${\bar W}$ or ${\bar F}_{\bar m}$, leading to the same phase $%
-\gamma_W$ as seen in (\ref{gauginoam}). Thus, the phase of the gauginos is $%
\phi_{M_a}=-\gamma_W$.

The Yukawa couplings also have phases in general. As mentioned in section %
\ref{G2MSSM}, in our analysis we have not attempted to explain the
origin of Yukawa couplings and have assumed that the Yukawa
couplings are the same as that in the Standard-Model. This
effectively means that the Yukawas just contain one non-trivial
phase, the one which is present in the CKM matrix.

The phases of supersymmetry breaking trilinear couplings can also be
computed in terms of $\gamma _{W}$. From (\ref{trilinear}), $A_{\alpha \beta
\gamma }\propto F^{\phi }\hat{K}_{\phi }=K^{\phi \bar{\phi}}F_{\bar{\phi}}%
\hat{K}_{\phi }$. Then one can show that ${\hat{K}}_{\phi }=\phi
_{0}e^{-i\theta }$ and $F_{\bar{\phi}}\sim e^{i\theta -i\gamma _{W}}$.
Therefore, the overall phase for the trilinear coupling is simply $\phi
_{A}=-\gamma _{W}$, the same as that for gaugino masses.

The $\mu$ and $B\mu$ parameters (with $\mu^{\prime }=0$ \footnote{%
which is favored by the motivation of the solution to the doublet-triplet
splitting problem\cite{Witten:2001bf}.}) can be written as \cite%
{Brignole:1997dp}:
\begin{eqnarray}
\mu&=&\Big(e^{i\gamma_W}m_{3/2}Z-e^{{\hat K}/2}F^{\bar m}\partial_{\bar m}Z%
\Big)({\tilde K}_{H_u}{\tilde K}_{H_d})^{-1/2}  \notag \\
B\mu&=& \bigg\{(2m_{3/2}^2 +V_0)Z- e^{i\gamma_W}m_{3/2} e^{{\hat
K}/2}F^{\bar
m}\partial_{\bar m}Z+e^{i\gamma_W}m_{3/2} e^{{\hat K}/2}F^m [\partial_m Z -Z\partial_m \ln(%
{\tilde K}_{H_u}{\tilde K}_{H_d})]  \notag \\
&-&e^{{\hat K}}F^{\bar m}F^{n}[\partial_{\bar m}\partial_n
Z-\partial_{\bar m}Z
\partial_n \ln({\tilde K}_{H_u}{\tilde K}_{H_d})]\bigg\}({\tilde K}_{H_u}{%
\tilde K}_{H_d})^{-1/2}
\end{eqnarray}
where $Z$ is the Higgs bilinear coefficient and $\tilde{K}$ is the K\"{a}%
hler metric for the Higgs fields. $Z$ is a complex-valued function of all
hidden sector chiral fields in general. Therefore, in general both $\mu$ and
$B\mu$ have independent phases. However, with the reasonable assumption that
the visible sector K\"{a}hler metric and the Higgs bilinear coefficient $Z$
are independent of the meson field $\phi$, one can make simplifications.
Combining the above with the fact that $F^i \ll F^{\phi}$, one has the
following approximation:
\begin{eqnarray} \label{mu}
\mu&\approx& e^{i\gamma_W}m_{3/2}Z\,({\tilde K}_{H_u}{\tilde K}%
_{H_d})^{-1/2}\equiv e^{i\gamma_W}m_{3/2}Z_{\mathrm{eff}}  \notag \\
B\mu&\approx& (2m_{3/2}^2 +V_0)Z\,({\tilde K}_{H_u}{\tilde K}%
_{H_d})^{-1/2}\equiv(2m_{3/2}^2 +V_0)Z_{\mathrm{eff}}
\end{eqnarray}
Therefore, the phases of $\mu$ and $B\mu$ are:
\begin{eqnarray}
\phi_{\mu}&\approx&\gamma_W+\gamma_Z  \notag \\
\phi_{B\mu}&\approx&\gamma_Z
\end{eqnarray}
To summarize, with some reasonable assumptions, at leading order
in our analysis, all soft terms depend on only two phases -
$\gamma_W$ and $\gamma_Z$. This means that the reparameterization
invariant phases given by $\phi_{\mu}+\phi_A-\phi_{B\mu}$ and
$\phi_{\mu}+\phi_{M_a}-\phi_{B\mu}$ \cite{Chung:2003fi} vanish,
implying that within the above assumptions there are no
non-trivial CP-violating phases beyond that in the Standard-Model
at the unification scale. It is possible to do a global phase
transformation of the superpotential without
affecting physical observables. Thus, one can choose a basis in which $%
\gamma_W$ vanishes. The reparameterization invariant combinations above
still give the same result.

RG evolution to low scales can lead to additional effects. There is a finite
threshold correction to the bino and wino mass parameters from the
Higgs-Higgsino loop at low-scales which is proportional to $\mu $ (more
about this in the next section). The correction thus depends on the phase $%
\phi _{\mu }$ making the phase of the low scale bino and wino mass
parameters different from $-\gamma _{W}$ in general. Therefore, the
reparameterization invariant combinations $\phi _{\mu }+\phi _{M_{a}}-\phi
_{\mu }$ for $a=1,2$ will not vanish in general at low scales giving rise to
non-trivial phases. In the basis in which $\gamma _{W}$ vanishes, these
non-trivial phases will depend on $\gamma _{Z}$. At present, it is not
possible to reliably compute $\gamma _{Z}$ from first principles. Therefore,
in our analysis below, we will only study situations with $\gamma _{Z}=0$ or
$\pi $ \footnote{%
This is in the basis in which $\gamma _{W}=0$.}. This corresponds to $\mu $
being positive or negative respectively. We hope to study non-trivial phases
in more detail in the future.

\subsection{Summary of the $G_2$-MSSM  at the Unification scale.}

In summary, the soft-susy breaking parameters at the unification
scale in the $G_2$-MSSM are given by equations $(20)$, $(23)$,
$(32)$ and $(36)$ (with $z_{eff}$ order one). The microscopic
constants which are determined by the $G_2$-manifold are subject
to the constraints discussed at the beginning of section II. We
now turn to the detailed discussion of renormalising the masses
and couplings down to the Electroweak scale.

\section{Superpartner Spectrum at $M_{EW}$ and Electroweak Symmetry Breaking}

\label{spectra}

As seen in previous section, the scalar and Higgsino masses at $M_{\mathrm{%
unif}}$ are close to that of the gravitino. This has to be larger than $%
\gtrsim 10~\mathrm{TeV}$ in order to evade the LEP II chargino bound because
of the large suppression of the gaugino masses relative to the gravitino. In
addition, a gravitino mass of $\gtrsim 10$ TeV is also required to mitigate
the moduli and gravitino problems.

In order to connect to low-energy physics, one has to RG evolve the soft
supersymmetry breaking parameters from $M_{\mathrm{unif}}$ to the
electroweak scale. It turns out that RG evolution increases the masses of
the first and second generation squarks and sleptons. However, since the
increase is mostly proportional to the gaugino masses \cite{Martin:1997ns}
which are much smaller than the high-scale sfermion mass, the masses of the
first and second generations squarks and sleptons are still of $\mathcal{O}%
(m_{3/2})$. The masses of the third generation squarks and sleptons - stops,
sbottoms and staus are also affected non-trivially by the trilinear
parameters (again of $\mathcal{O}(m_{3/2})$) because of their larger Yukawa
couplings \cite{Martin:1997ns}. In particular, the lightest stop $(\tilde{t}%
_1)$ becomes much lighter than the other sfermions (even though still
considerably heavier than the gauginos). Finally, the $\mu$ parameter, which
determines the masses of the Higgsinos, does not change much during RG
evolution because of the non-renormalization theorem\footnote{%
it only suffers from wave-function renormalization.}. So, the $\mu$
parameter at the electroweak scale is also of $\mathcal{O}(m_{3/2})$.

Because of the large hierarchy in the spectrum, it is convenient to work in
an effective theory with the heavy fields (scalars and Higgsinos) integrated
out at their characteristic scale ($M_s \sim 10-100$ TeV). The low energy
effective theory below $M_s$ only contains the light gauginos and the SM
particles. Therefore it is very important to compute the masses of gauginos
as these are the only light new states predicted by the theory\footnote{%
except possibly the lightest stop.}. To take into account the threshold
effects of these heavy states on the gaugino masses, we follow the `match
and run' procedure which is a good approximation when $M_a \ll M_s$. In this
paper, we use a one-loop two-stage RGE running with a tree-level matching at
the scale $M_s$. All other thresholds are calculated using the exact
one-loop results.

\subsection{Gaugino Masses at $M_{EW}$}

\label{gauginospec}

The weak scale gaugino mass parameters at one-loop can be related to those
at unification scale by a RG evolution factor $K_a$ as follows:
\begin{eqnarray}  \label{gauginolow}
M_a(M_{\mathrm{weak}}) &=& K_a\,M_a(M_{\mathrm{unif}})
\end{eqnarray}
The RG evolution factors $K_a$ are given by:
\begin{eqnarray}
K_a = \left(\frac{\alpha_a^{s}}{\alpha_{ unif}}\right)\left(\frac{%
\alpha_a^{EW}}{\alpha_a^{s}}\right)^{{\tilde b}_a^{SM}/b_a^{SM}}
\end{eqnarray}
where ${\tilde b}_a$'s and $b_a$'s are the $\beta$ function coefficients of
the gaugino masses and gauge couplings respectively:
\begin{eqnarray}
{\tilde b}_1^{SM} &=& 0, \quad {\tilde b}_2^{SM} = -6, \quad {\tilde b}%
_3^{SM} = -9  \notag \\
b_1^{SM} &=& \frac{41}{10}, \quad b_2^{SM}= -\frac{11}{6}, \quad b_3^{SM}=-5.
\end{eqnarray}
$\alpha_a^s$ and $\alpha_a^{EW}$ are the gauge couplings at the decoupling
scale $M_s$ and the electro-weak scale $M_{\mathrm{EW}}$ respectively, which
can be expressed as
\begin{eqnarray}
(\alpha_a^{s})^{-1}&=&\alpha_{ unif}^{-1}+\frac{b_a}{2\pi}\ln\left(\frac{M_{%
\mathrm{unif}}}{M_s}\right)  \notag \\
(\alpha_a^{EW})^{-1}&=&(\alpha_a^{s})^{-1}+\frac{b_a^{SM}}{2\pi}\ln\left(%
\frac{M_s}{M_{ EW}}\right)
\end{eqnarray}
As an example, for $\alpha_{unif}=1/26.5$, for $M_s$ varying from $10~%
\mathrm{TeV}$ to $100~\mathrm{TeV}$, the corresponding RG factors are
\begin{eqnarray}  \label{rg-correc}
K_1\approx 0.47 - 0.49, \quad K_2\approx 0.99 - 1.08, \quad K_3\approx 3.7 -
4.6.
\end{eqnarray}
Notice that the RG evolution factor $K_3$ is quite large for a large $M_s$.
This prevents the gluino becoming the LSP even though the gluino mass
$M_3$ is typically small at the unification scale.

Once the running masses of the gauginos at the low scale are calculated,
their pole mass can be obtained by adding weak scale threshold corrections.
For general MSSM parameters, they are given in \cite{BMPZ}. In our `match
and run' procedure, the threshold corrections of heavy scalars and Higgsinos
are automatically included except some finite terms which are usually small
and negligible. However, since in our case the Higgsino mass $\mu$ is of
order $m_{3/2}$, the finite threshold correction cannot be neglected and is
given by \cite{BMPZ,chisplit3,ADG}:
\begin{eqnarray}  \label{Higgs-Higgsino}
\Delta M_{1,2}^{\mathrm{finite}}&\approx & -\frac{\alpha_{1, 2}}{4\pi}\,%
\frac{\mu\, \sin{(2\beta)}}{(1-\frac{\mu^2}{m_A^2})}\ln(\frac{\mu^2}{m_A^2})
\notag \\
&\approx& \frac{\alpha_{1,2}}{4\pi}\,\mu = \frac{\alpha_{1,2}}{4\pi}\,z_{%
\mathrm{eff}}\;m_{3/2}  \label{Eq:gaugino-correction}
\end{eqnarray}
In the second line, we have used the fact that $\frac{\mu^2}{m_A^2}\sim 1$
so that the logarithm can be expanded. In addition, since $\mu$ does not
change much in the RG evolution, it is a good approximation to use its high
scale value $\mu\equiv z_{\mathrm{eff}}\,m_{3/2}$. This quantum correction
proportional to $\mu$ will shift the bino ($M_1$) and wino ($M_2$) masses up
or down depending on the sign of $\mu$. This will most significantly affect $%
M_2$ as it is typically the lightest gaugino at $M_{\mathrm{unif}}$.
Therefore it could potentially affect the identity of the LSP. For the case
with gravitino mass $m_{3/2}\sim 30 ~\mathrm{TeV}$, this finite correction
to $M_2$ is roughly $100$ GeV, which may even dominate over the tree-level
mass. This large correction to $M_{1,2}$ is not surprising since the low
energy effective theory is non-supersymmetric and there is no symmetry to
protect the gaugino masses from finite radiative corrections.

\begin{figure}[h!]
\begin{tabular}{c}
\leavevmode \epsfxsize 14.5 cm \epsfbox{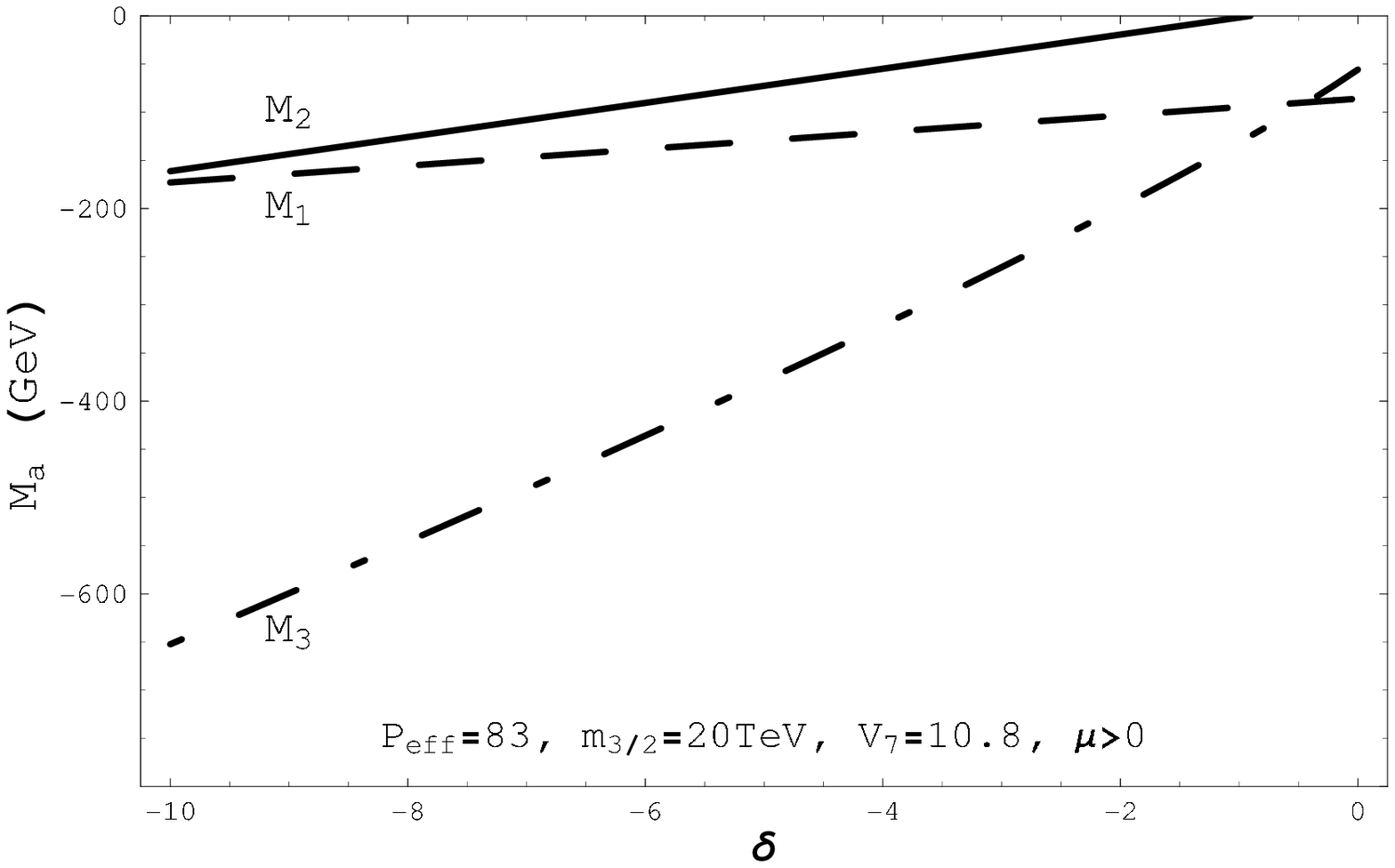} \\
\leavevmode \epsfxsize 14.5 cm \epsfbox{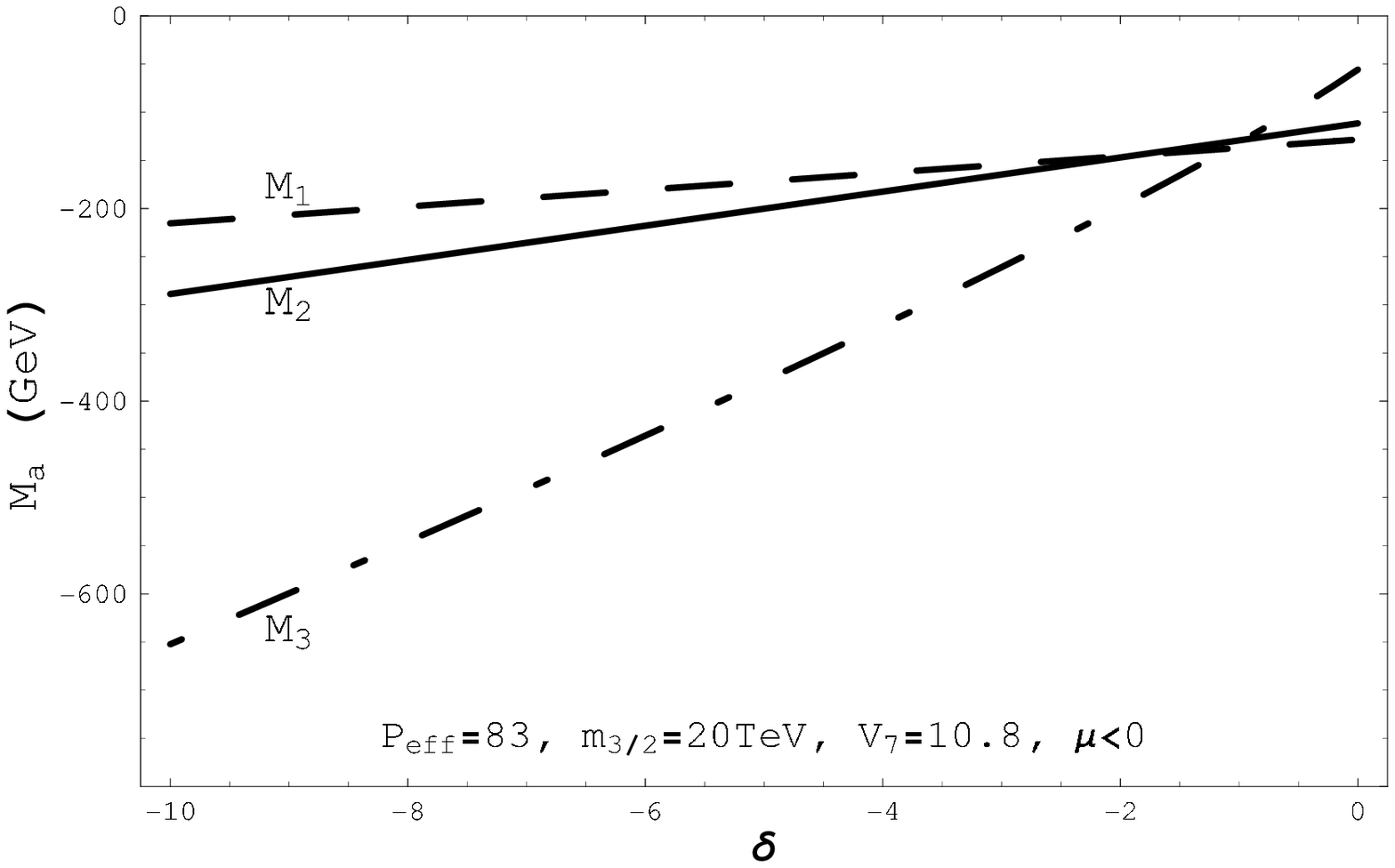} \\
\end{tabular}
\caption{{\protect\footnotesize Gaugino Masses at low scales (include
correction (\protect\ref{Eq:gaugino-correction})) as a function of $\protect%
\delta$ for the case with cosmological constant tuned to zero ($P_{\mathrm{%
eff}} = 83$), $Q-P=3$, $V_7=10.8$, $C_2=1$ and $\protect\alpha%
_{unif}^{-1}=26.5$.TOP: Plot for positive $\protect\mu$. BOTTOM: Plot for
negative $\protect\mu$.}}
\label{Fig:gaugino-low1}
\end{figure}

It is important to notice that since the above correction is linear in $\mu$%
, it is \emph{sensitive} to the sign of $\mu$. More generally, as was
explained in section \ref{phases}, $\mu$ can have a different non-trivial
phase from the gaugino masses which share a common phase. So the inclusion
of the finite threshold correction can eventually lead to non-trivial phases
for the reparameterization invariant combinations. This can change the
spectrum eigenvalues, leading to observable effects for colliders, for
Dark Matter relic density, Dark Matter detection and for Electric Dipole
Moments (EDMs). As mentioned in section \ref{phases}, it is unfortunately
not possible at present to compute the effects of these phases reliably, so
we will only consider the effects of positive and negative $\mu$
(corresponding to $\gamma_Z=0$ and $\pi$ in our analysis respectively%
\footnote{%
This is in the basis in which $\gamma_W=0$.}). This leads to two different
plots as shown in Figure \ref{Fig:gaugino-low1}.

Of course, one also has to include weak scale threshold corrections from
gaugino-gauge-boson loops. These are especially important for the gluino
mass:
\begin{eqnarray}
\Delta M_3^{\mathrm{rad}}=\frac{3g_3^2}{16\pi^2}\left(3\ln\left(\frac{M_{%
\mathrm{EW}}^2}{M_3^2}\right)+5\right)M_3
\end{eqnarray}
For $M_{3}$ not much heavier than $M_{\mathrm{EW}}$, there is a substantial
correction of at least $3\alpha_3 M_3$.

We calculated the gaugino mass at the Weak scale, with all these corrections taken into
account. As mentioned before, the hierarchy of gaugino masses is most
sensitive to $\delta$ and $P_{\mathrm{eff}}$. In order to obtain realistic
phenomenology, we choose $Q-P=3$ and tune the cosmological constant to
obtain $P_{\mathrm{eff}}=83$. Then, the hierarchy of gaugino masses mainly
depends on $\delta$ - the threshold corrections to the unified gauge
coupling. The dependence of gaugino masses on $\delta$ is shown in Figure %
\ref{Fig:gaugino-low1}. From the figures, we see that for $\mu>0$, the wino
tends to be the LSP ($|M_2|<|M_1|$) while for $\mu<0$, the bino tends to be
the LSP ($|M_1|<|M_2|$) for a large range in $\delta$. Also, $|M_3|$ is
significantly larger than $|M_1|,|M_2|$ for values of $|\delta|\gtrsim 3$.
As will be seen in section \ref{gaugino-gcu}, $|\delta|\gtrsim 3$ is favored
by precision gauge unification.

\subsection{Electroweak Symmetry Breaking}

Since the scalars in the $G_2$-MSSM are generically very heavy, there are
large logarithmic corrections to the Higgs potential. To analyze EWSB in
such a theory, it is better to work in the low energy effective field theory
in which all heavy fields are decoupled. Then, the large logarithmic
corrections are automatically resummed when the Higgs parameters are RG
evolved to the low scale. After decoupling all the heavy fields, the low
energy effective theory is simply the standard model plus light gauginos. It
is well known that in order to have electroweak symmetry breaking, there
must exist a light Higgs doublet below the decoupling scale with negative
mass parameter. It was pointed out in \cite{DG} that generically it is very
hard to get EWSB predominantly from radiative effects below the decoupling
scale, more so if the decoupling scale is not too high (as in our case).
Therefore, if electroweak symmetry breaking happens in the effective theory
at low scale, it should also happen in the MSSM theory at the decoupling
scale. %Therefore if a theory has electroweak symmetry breaking, it should
%be independent of the scale of the analysis, whether it is done in
%the weak scale effective theory or in the MSSM at the decoupling
%scale.
This means that we only need to check the existence of EWSB at the
decoupling scale in the MSSM framework.

%So we only need to check whether there is a small($\sim M_{EW}$)
%negative eigenvalue at decoupling scale after we diagonalize the
%mass matrix of the two Higgs bosons. If the EWSB is generic in the
%parameter space we have scanned, a light eigenvalue should occur
%very often.
In order to do that, we have to first diagonalize the mass matrix of $%
(h_u,h_d^*)$:
\begin{eqnarray}
\left(
\begin{array}{cc}
m_{H_u}^2+\mu^2 & -b \\
-b & m_{H_d}^2+\mu^2%
\end{array}
\right)  \label{Eq-Higgsmass}
\end{eqnarray}
The eigenvalues are
\begin{eqnarray}
\zeta_{1,2} = \frac{1}{2}\Big [(m_u^2+m_d^2)\pm\sqrt{(m_u^2-m_d^2)^2+4b^2}%
\Big],
\end{eqnarray}
where $m_u^2=m_{H_u}^2+\mu^2$ and $m_d^2=m_{H_d}^2+\mu^2$. The light Higgs
doublet $h$ is a superposition of $h_u$ and $h_d$
\begin{eqnarray}
h=\sin\beta h_u +\cos\beta h_d^*
\end{eqnarray}
where $\beta$ is determined by the diagonalization of the matrix. In a
complete high scale theory, the mass matrix Eq.(\ref{Eq-Higgsmass}) is
completely determined by the high scale boundary condition. The existence of
EWSB depends on whether there is one negative eigenvalue. As explained
earlier, in realistic $G_2$ theory vacua, the effective $\mu$ and $B\mu$
terms in the low-energy lagrangian arise from the non-zero Higgs bilinear
coupling $Z$. If the $\mu$ term in the original superpotential is forbidden
by some discrete symmetry as in \cite{Witten:2001bf}, then they are given by
(see section \ref{phases}):
\begin{eqnarray}
\mu(M_{\mathrm{unif}}) \approx z_{\mathrm{eff}}m_{3/2}, \quad B\mu(M_{%
\mathrm{unif}}) \approx 2z_{\mathrm{eff}}m_{3/2}^2
\end{eqnarray}
Using the above relation and RG evolving them to the decoupling scale, one
can obtain a Higgs mass matrix which is parameterized by $z_{\mathrm{eff}}$.
One then finds that for $z_{\mathrm{eff}} < z_{\mathrm{eff}}^*\sim \mathcal{O%
}(1)$ there will always be a negative eigenvalue and so electroweak symmetry
is broken. This condition for the existence of EWSB is naturally satisfied
if $z_{\mathrm{eff}} \sim \mathcal{O}(1)$. Since all elements in the mass
matrix (\ref{Eq-Higgsmass}) are $\mathcal{O}(m_{3/2})$, the mixing
coefficients ($\sin\beta$ and $\cos\beta$) of $h_u$ and $h_d^*$ are of the
same order. Thus, $\tan\beta$ is naturally predicted to be of $\mathcal{O}(1)
$. This is in contrast to usual approaches to high-scale model-building
where $\mu$ and $B\mu$ are completely unknown from theory and are only
determined after fixing $M_Z$ and choosing $\tan(\beta)$. In the $G_2$-MSSM,
$\tan\beta$ is not a free parameter and is determined by the relation
between $\mu$ and $B\mu$ predicted from the theory.

Since all the elements in the mass matrix (\ref{Eq-Higgsmass}) are $\mathcal{%
O}(m_{3/2})$, the Higgs mass eigenvalues should also be $m_{3/2}$ which is
around 100 times larger than the EW scale. This implies a fine-tuning if
there is no magic cancellation. So, even though the existence of EWSB is
generic, getting the correct $Z$ mass is not. As will be discussed below,
the requirement of obtaining the correct $Z$-boson mass fixes the precise
value of $z_{\mathrm{eff}}$. This requires a fine-tuning of $z_{\mathrm{eff}}
$.

In the following we describe the precise procedure used for
obtaining EWSB with the correct $Z$ mass. The heavy scalars are
decoupled at $M_{s}$ and the couplings of the low energy effective
theory (consisting of the SM particles and the MSSM gauginos) are
matched with those of the complete MSSM. Most importantly, the
matching condition for the quartic coupling of the Higgs is given
by:
\begin{equation}\label{lambda}
\lambda (m_{s})=\frac{\frac{3}{5}g_{1}^{2}+g_{2}^{2}}{8}\cos
^{2}2\beta
\end{equation}
It turns out that at energies below the decoupling scale $M_{s}$, the
\textit{one-loop} RG evolution of $m$ (the SM Higgs mass parameter), $%
\lambda $ and the Yukawa couplings is the same as that of the Standard
Model:
\begin{eqnarray}\label{RGE}
16\pi ^{2}\frac{d\lambda }{dt} &=&24\lambda
^{2}-6y_{t}^{4}+12\lambda
y_{t}^{2}+\frac{27}{200}g_{1}^{4}+\frac{9}{20}g_{1}^{2}g_{2}^{2}+\frac{9}{8}%
g_{2}^{4}-\frac{9}{5}\lambda g_{1}^{2}-9\lambda g_{2}^{2} \nonumber\\
16\pi ^{2}\frac{dm^{2}}{dt} &=&m^{2}(6\lambda +6y_{t}^{2}-\frac{9}{10}%
g_{1}^{2}-\frac{9}{2}g_{2}^{2}) \nonumber\\
16\pi ^{2}\frac{dy_{t}}{dt} &=&y_{t}\Big[(\frac{9}{2}y_{t}^{2}+\frac{3}{2}%
y_{b}^{2}+y_{\tau }^{2})-(\frac{17}{20}g_{1}^{2}+\frac{9}{4}%
g_{2}^{2}+8g_{3}^{2})\Big] \nonumber\\
16\pi ^{2}\frac{dy_{b}}{dt} &=&y_{t}\Big[(\frac{3}{2}y_{t}^{2}+\frac{9}{2}%
y_{b}^{2}+y_{\tau }^{2})-(\frac{1}{4}g_{1}^{2}+\frac{9}{4}%
g_{2}^{2}+8g_{3}^{2})\Big] \nonumber\\
16\pi ^{2}\frac{dy_{\tau }}{dt} &=&y_{\tau }\Big[(3y_{t}^{2}+3y_{b}^{2}+%
\frac{5}{2}y_{\tau }^{2})-(\frac{9}{4}g_{1}^{2}+\frac{9}{4}g_{2}^{2})\Big]
\end{eqnarray}
It is important to mention that the above equations are different from the
corresponding ones for split-supersymmetry because Higgsinos in the $G_{2}$%
-MSSM are also very heavy (of $\mathcal{O}(m_{3/2})$) unlike that in
split-supersymmetry \cite{ADGW}. Therefore, for the $G_{2}$-MSSM, the
Higgsinos also decouple below $M_{s}$ forbidding additional terms which
appear in the one-loop RG equations for split-supersymmetry. From above, the
quartic coupling $\lambda $ will get a large correction from RG evolution
because of the large top Yukawa coupling. The gaugino masses $M_{a}$ on the
other hand will only receive one-loop corrections from gauge boson exchange.
The corresponding RGE equations at one-loop are as follows:
\begin{eqnarray}
16\pi ^{2}\frac{dM_{1}}{dt} &=&0 \\
16\pi ^{2}\frac{dM_{2}}{dt} &=&-12b_{2}^{2}M_{2} \\
16\pi ^{2}\frac{dM_{3}}{dt} &=&-18b_{3}^{2}M_{3}
\end{eqnarray}%
Given the boundary conditions for soft parameters for realistic $M$ theory
vacua as in section \ref{soft-unif}, one finds that EWSB occurs if $z_{%
\mathrm{eff}}$ is of $\mathcal{O}(1)$. But the generic value of $M_{Z}$ is
around $m_{3/2}$, which can be seen from the fact that all the Higgs
parameters are of the order of $m_{3/2}$. In order to get $M_{Z}=91~\mathrm{%
GeV}$, one has to tune $z_{\mathrm{eff}}$ so that $\mu $ and $B\mu $ take
values such that the lightest Higgs mass parameter comes out to be around $%
M_{EW}$. This fine-tuning is a manifestation of the \emph{little hierarchy
problem} - an unexplained hierarchy between the electroweak (SM-like Higgs)
and superpartner (scalar) scales. Our current understanding of the theory
does not yet allow us to explain the little hierarchy problem by a dynamical
mechanism.

%We start with our high scale boundary conditions obtained from
%M-theory model. Since the gaugino masses is generically suppressed
%at least ${\cal O}(10)$, their contributions to the RGE running of
%scalar masses and trilinears can be neglected. In addition $\mu$
%and $B\mu$ parameters evolve independent of other scalar masses
%and trilinears. So varying $z_{\rm eff}$ at the boundary condition is
%equivalent to varying the mass parameter of the light Higgs
%doublet. In order to achieve small $m_H^2$, we only need to tune
%$Z_{\rm eff}$.

%The analytic result for low scale parameters in terms of the high
%scale parameters.\\
%.........

\begin{figure}[h]
\epsfxsize 14 cm \epsfbox{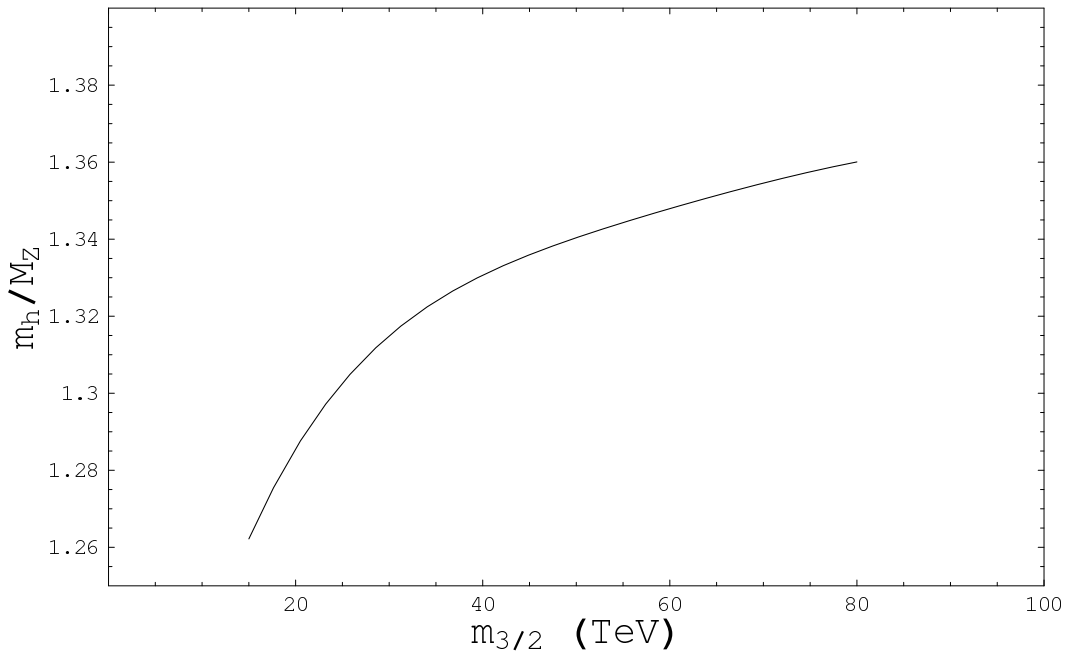} \caption{Plot of the ratio of
the Higgs mass and the $Z$ mass as a function of the gravitino
mass.} \label{mhmz}
\end{figure}

In the low energy effective theory the ratio of the Higgs mass to
the $Z$ mass turns out to be quite robust\footnote{even when one
does not tune $z_{eff}$ to obtain the correct $Z$ mass as
explained in the previous paragraph.}. The ratio is given by: \ba
\frac{m_h}{m_Z}=\frac{2\sqrt{\lambda}
(M_{EW})}{(\frac{3}{5}g_1^2+g_2^2)(M_{EW})}\ea $\lambda$ is
determined by the gauge couplings, Yukawa couplings and
$\tan\beta$. One has to use the boundary condition for $\lambda$
at $M_s$ as in (\ref{lambda}) and then RG evolve it to the
electroweak scale using the first equation in (\ref{RGE}). The
gauge couplings at the electroweak scale can also be determined by
their RGEs. Thus, one can obtain the ratio $m_h/m_Z$ as a function
of $m_{3/2}$ as shown in Figure \ref{mhmz}. It is worth noting
that this ratio only mildly depends on $m_{3/2}$.
If $M_Z$ is tuned to its experimental value, we can use $\lambda$
obtained from the RG equation or from Figure \ref{mhmz} to predict
the Higgs boson mass for
any given value of $m_{3/2}$. Once one finds the Higgs VEV $v$, the Higgs mass is simply $%
m_h^2=2\lambda v^2$ just as in the Standard Model since all heavy
scalars and Higgsinos have already been decoupled. The Higgs boson
mass thus computed turns out to be of $\mathcal{O}(120)$ GeV for a
range of interesting values of $m_{3/2}$ as it only mildly depends
on it. Since all susy-breaking large logarithms have already been
taken into account in the `Match and Run' procedure, only some
finite term contributions could have been missed in this analysis.
One could take those effects into account as well in a more
detailed analysis, but that would not change the Higgs mass
significantly. The origin of the above value of the Higgs mass can
be understood as follows. If one did not decouple the scalars and
Higgsinos at $M_s$, then the Higgs mass receives very large
radiative corrections, making the Higgs mass heavy as required.
However, because of the hierarchy between the scalars and
gauginos, these radiative corrections are hard to compute in a
controlled manner. In the spirit of effective field theory
therefore, it makes sense to integrate out the scalars and
Higgsinos at $M_s$. In this picture, all the radiative corrections
to the Higgs mass can be incorporated in the running of the
quartic coupling $\lambda$. $\lambda$ gets renormalized from $M_s$
to $M_{EW}$ giving rise to a heavy Higgs.

\section{Gauge Coupling Unification}

\label{gaugino-gcu}

In section \ref{G2MSSM}, it was mentioned that in many cases the strong
coupling limit of $E_8 \times E_8$ heterotic string theory compactifications
on a Calabi-Yau threefold $Z$ is the same as $M$ theory compactifications on
a singular $G_2$-holonomy manifold $X$. Since a GUT-like spectrum is natural
in weakly coupled heterotic compactifications, a GUT-like spectrum (breaking
down to the MSSM by Wilson-lines) was assumed for $G_2$ compactifications in
our study as well. At a theoretical level, because of an underlying GUT
structure, the MSSM gauge couplings are unified at the compactification
scale $M_{KK}$ for both heterotic and $G_2$ compactifications. However, when
one tries to impose constraints from the extrapolated values of observed
gauge couplings, interesting differences arise between weakly coupled
heterotic and $G_2$ compactifications. Here, we will first explain the
difference between weakly coupled heterotic compactifications and $G_2$
compactifications regarding gauge unification and then discuss the procedure
used in our analysis to obtain sets of parameters compatible with \emph{%
precise} gauge unification.

In weakly coupled heterotic string compactifications, there is a relation
between the Newton's constant $G_N$, the unified gauge coupling $\alpha_{%
\mathrm{unif}}$, the string coupling $e^{\phi}$ and the volume of the
internal manifold $V_{CY}$ \cite{Witten:1996mz}. Assuming a more or less
isotropic Calabi-Yau, one has $V_{CY} \sim M_{\mathrm{unif}}^{-6}$ which
gives:
\begin{eqnarray}
G_N \approx \frac{\alpha_{\mathrm{unif}}^{4/3}}{4M_{\mathrm{unif}}^2}\left(%
\frac{16\pi}{e^{2\phi}}\right)^{1/3} > \frac{\alpha_{\mathrm{unif}}^{4/3}}{%
4M_{\mathrm{unif}}^2}
\end{eqnarray}
since the string coupling is weak by assumption $(e^{2\phi}<1)$.
Substituting the values of $\alpha_{\mathrm{unif}}$ and $M_{\mathrm{unif}}$
obtained by extrapolating the observed gauge couplings in the MSSM, the
prediction for $G_N$ turns out to be too large compared to the observed
value. Various proposals have been put forward for dealing with this problem
within the perturbative heterotic setup, but none of them are obviously
compelling.

\begin{figure}[h]
%\resizebox{100mm}{!}{\includegraphics{Lplot.eps}} \leavevmode
\epsfxsize 14 cm \epsfbox{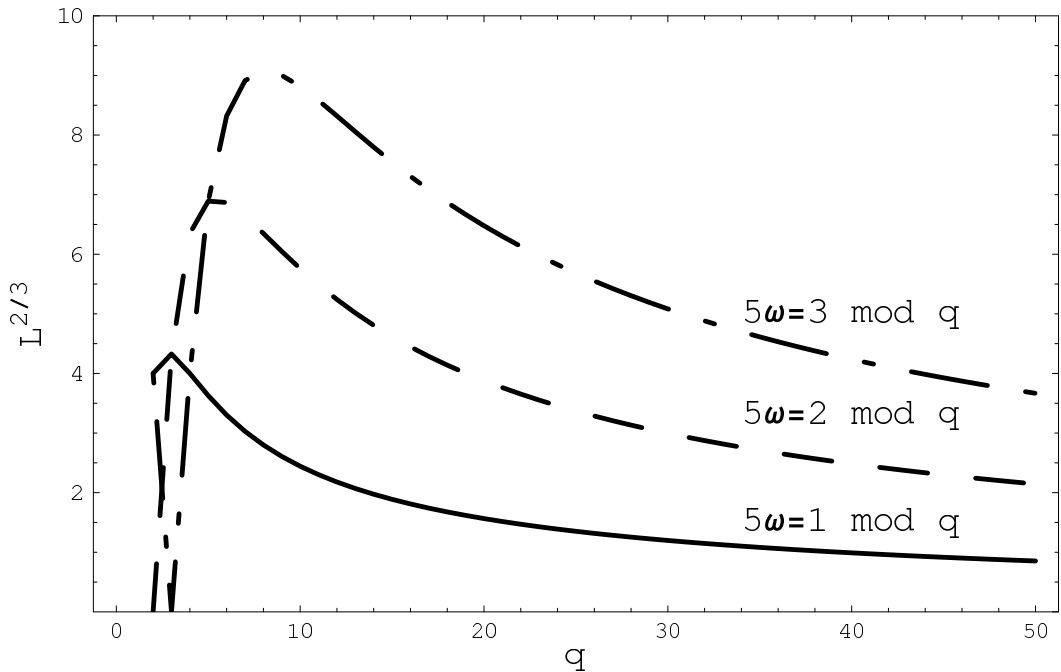}
\caption{Plot of $L^{2/3}$ as a function of integer $q$ from $2$ to $50$.
Different curves correspond to different choices of $\protect\omega $ as
marked in the plot. }
\label{fig:Lplot}
\end{figure}

In $G_{2}$ compactifications however, one finds a different relation among
the same quantities\footnote{%
in $M$ theory however, there is no string coupling.} after doing a similar
analysis \cite{Friedmann:2002ty}:
\begin{equation}
G_{N}\approx \frac{\alpha _{\mathrm{unif}}^{2}}{32\pi ^{2}M_{\mathrm{unif}%
}^{2}}\left( \frac{1}{a}\right) ;\;\;a\equiv \frac{V_{7}}{V_{\hat{Q}%
_{vis}}^{7/3}}
\end{equation}
Here, \textquotedblleft $a$" is the dimensionless ratio between the volume
of the $G_{2}$ manifold $V_{7}$ and the volume of the three-manifold $\hat{Q}%
_{vis}$ on which the visible sector MSSM gauge group is supported. If one
does a more careful analysis and takes into account the threshold
corrections to the unified gauge coupling from Kaluza-Klein (KK) modes, one
obtains \cite{Friedmann:2002ty}:
\begin{equation}\label{correct}
G_{N}=\frac{\alpha _{\mathrm{unif}}^{2}}{32\pi ^{2}M_{\mathrm{unif}}^{2}}%
\left( \frac{(L(\hat{Q}_{vis}))^{2/3}}{a}\right)
\end{equation}
where $L(\hat{Q}_{vis})$ is the contribution from the threshold correction.
Substituting the values of $\alpha _{\mathrm{unif}}$ and $M_{\mathrm{unif}}$
obtained by extrapolating the observed gauge couplings in the MSSM and the
value of $G_{N}$, one finds:
\begin{equation}\label{constraint}
\left( \frac{(L(\hat{Q}_{vis}))^{2/3}}{a}\right) \approx 15
\end{equation}
In all examples where duality with heterotic string theory or Type
IIA string theory is used to deduce the existence of the $G_{2}$
manifold $X$,
\textquotedblleft $a$" is expected to be much less than unity \cite%
{Friedmann:2002ty}. For $G_{2}$-MSSM vacua, with natural values of
the microscopic parameters one obtains $V_{7}=10-100$
\cite{Acharya:2007rc} while $V_{\hat{Q}_{vis}}\sim \alpha
_{\mathrm{unif}}^{-1}\sim \mathcal{O}(25) $\footnote{
The unified coupling constraint will be discussed in Appendix \ref%
{constraints}.}. Thus, $a\equiv \frac{V_{7}}{V_{\hat{Q}_{vis}}^{7/3}}\ll 1$
is also naturally satisfied for $G_{2}$-MSSM vacua. In addition, by
expressing $V_{7}$ and $V_{\hat{Q}_{vis}}$ in terms of $M_{11}$ and $M_{%
\mathrm{unif}}$ respectively, $a<1$ implies $M_{\mathrm{unif}}<M_{11}$ which
means that the unification scale constraint stated in section \ref{soft-unif}
can be naturally satisfied. Since $a\ll 1$, from (\ref{constraint}) one
requires:
\begin{equation}
(L(\hat{Q}_{vis}))^{2/3}\ll 15
\end{equation}
The quantity $L(\hat{Q}_{vis}))^{2/3}$ depends on certain topological
invariants of the three-manifold $\hat{Q}_{vis}$ and can be computed for
special choices of $\hat{Q}_{vis}$. For one such choice - $\hat{Q}_{vis}=%
\mathbf{S^{3}}/Z_{q};\;q\in \mathbf{Z}$,
$(L(\hat{Q}_{vis}))^{2/3}$ has been computed
\cite{Friedmann:2002ty}. It depends on two integers $\omega ,q$
such that $5\omega $ is not a multiple of $q$.\footnote{It is
assumed that an $SU(5)$ GUT group is broken to the MSSM by Wilson
lines.} Figure \ref{fig:Lplot} shows the variation of $L^{2/3}$ for $\hat{Q}%
_{vis}=\mathbf{S^{3}}/Z_{q}$ as a function of $q$ for different choices of $%
5\omega $ mod $q$. One sees that $L^{2/3}\ll 15$ can be obtained in a
natural manner for a large range of $q$. For other choices of $\hat{Q}_{vis}$%
, it is reasonable to expect a similar result. To summarize therefore, $G_{2}
$-MSSM vacua are naturally compatible with gauge unification in general and
the \textquotedblleft unification scale constraint" mentioned in section \ref%
{soft-unif} in particular.

The value of the unified gauge coupling $\alpha_{\mathrm{unif}}$ is also
affected by the threshold corrections. The tree-level unified gauge coupling
of the visible sector at the compactification scale is the volume of the
visible sector three-manifold $\hat{Q}_{vis}$:
\begin{eqnarray}
\alpha_{vis}^{-1}\equiv \mathrm{Im}(f_{vis}^0) = V_{\hat{Q}_{vis}}=
\sum_{i=1}^{N} N_i^{vis} s_i,  \label{Eq:ineq2}
\end{eqnarray}
After taking into account the threshold corrections (at one-loop), one has:
\begin{eqnarray}  \label{gauge-coupling}
\alpha_{\mathrm{unif}}^{-1}&=&\alpha_{vis}^{-1}+\delta
\end{eqnarray}
For the one-loop result to be reliable, $\delta$ should be small compared to
$\alpha_{vis}^{-1}$. Since for the MSSM $\alpha_{unif}^{-1} \sim 25$ one
requires $\alpha_{vis}^{-1} \sim \mathcal{O}(25)$ as well. The conditions
under which the microscopic parameters can give rise to the above value of $%
\alpha_{vis}^{-1}$ is discussed in Appendix \ref{constraints}. The threshold
correction $\delta$ can be computed from the topological invariants of the
three-manifold $\hat{Q}_{vis}$, so it also depends on integers
characterizing the topology of $\hat{Q}_{vis}$. However, it is in general
regularization dependent in contrast to expression (\ref{correct}) for the
unification scale \cite{Friedmann:2002ty}. Therefore, in our analysis we
will assume $\delta$ to be a free parameter varying in a reasonable range
such that one-loop results are still reliable.

\subsection{Precision Gauge Unification}

Having convinced ourselves that $G_2$-MSSM vacua are naturally compatible
with gauge coupling unification, we will now examine the issue of precision
gauge coupling unification in the $G_2$-MSSM in the sense that we would like
to find sets of microscopic parameters which give rise to precise gauge
unification. There are two aspects to this issue: a) Is there a unification
of gauge couplings at a high scale $\mathcal{O}((1-3)\times 10^{16})~\mathrm{%
GeV}$ by continuing the gauge couplings up in energy from the
laboratory scale including all the low-scale thresholds, and b)
Whether the unified gauge coupling and the unification scale
obtained are consistent with the theoretical prediction in terms
of ``microscopic" parameters. As we will see, it turns out that
the $G_2$-MSSM is compatible with precision gauge coupling
unification in the sense that there exists a relatively large set
of reasonable microscopic parameters which gives rise to precise
gauge coupling unification.

Before going into details, it is important to notice the following general
fact -- gaugino masses at the unification scale and hence the low scale
depend on the value of $\alpha _{unif}^{-1}$. However, the value of $\alpha
_{unif}^{-1}$ itself depends on corrections to the gauge couplings from
superpartner thresholds at low scale. This means that there is a feedback
between the spectrum of superpartner masses and the value of $\alpha
_{unif}^{-1}$. Therefore, one has to remember to take into account the
effects of this feedback in general.
\begin{figure}[h]
\begin{tabular}{c}
\leavevmode \epsfxsize 17 cm \epsfbox{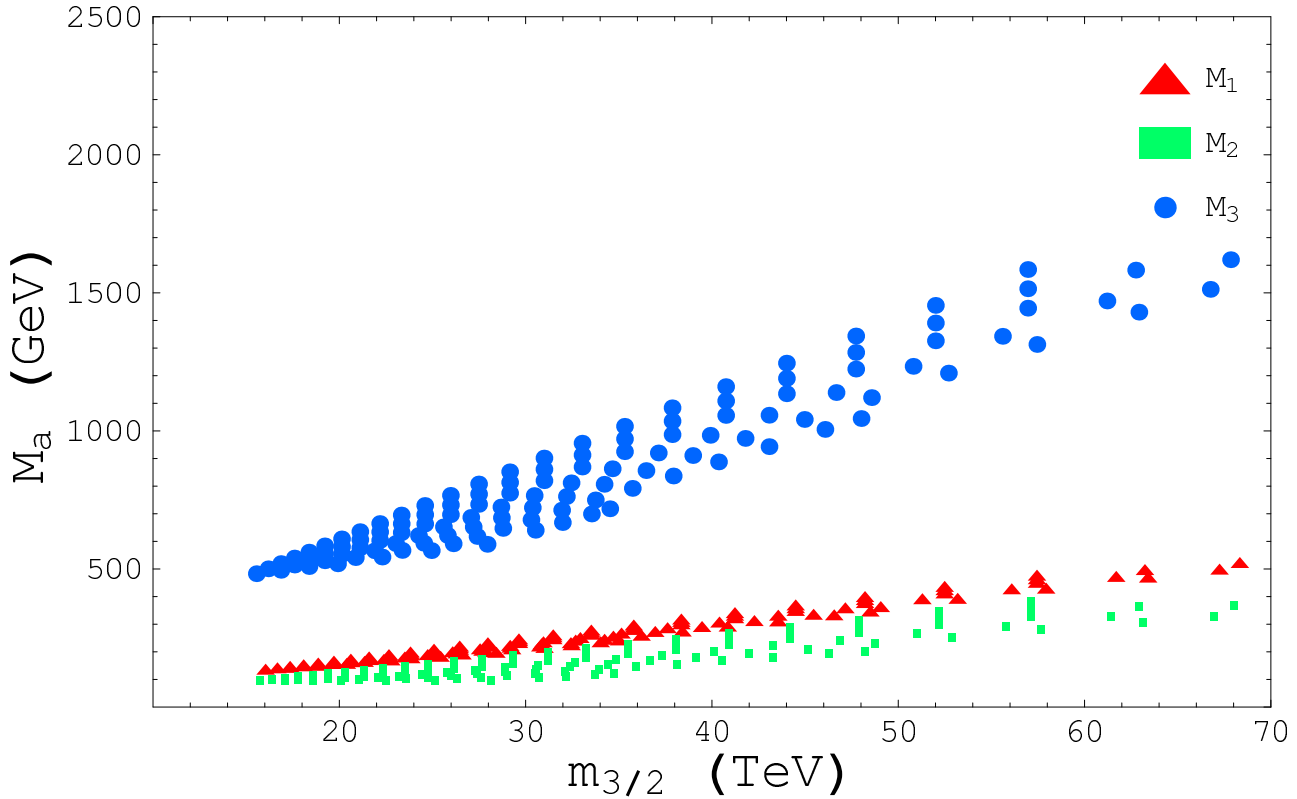}%
\end{tabular}
\caption{{\protect\footnotesize Gaugino Mass spectra vs $m_{3/2}$ compatible
with gauge unification for $P_{\mathrm{eff}}=83$, $C_{2}=5$ and $\protect\mu %
>0$. The red, green and blue lines correspond to gaugino mass $M_{1}$, $M_{2}
$ and $M_{3}$ respectively.}}
\label{Fig:scangcu1}
\end{figure}
Since the squarks and sleptons come in complete GUT multiplets, they do not
affect gauge unification. The Higgs doublets, Higgsinos and gauginos do
affect gauge unification since they do not form complete GUT multiplets.
Since the Higgsino mass ($\mu $) in these vacua are heavy ($\mathcal{O}%
(m_{3/2})$) and is robustly determined by the gravitino mass once the EWSB
breaking constraint is imposed, for a fixed gravitino mass gauge coupling
unification will mostly depend on the light gaugino masses, the gaugino mass
ratio $|M_{3}|/|M_{2}|$ in particular because it contributes the most to the
threshold corrections to the gauge couplings. We find that in order to have
precise unification, this ratio has to be greater than around $3-4$. This
sensitivity to $|M_{3}|/|M_{2}|$ is much greater here than in split-SUSY
where both the Higgsinos and gauginos are light. Finally, if there are
particles in the GUT multiplet in addition to the MSSM (the Higgs triplets
in the $SU(5)$ case for example) that are lighter than $M_{\mathrm{unif}}$,
then one should also take their threshold contributions to the gauge
couplings into account. However one finds that their threshold contribution
causes $\alpha _{3}^{-1}$ and $\alpha _{1}^{-1}$ to move \emph{away} from
each other. Therefore, the requirement of precision gauge unification forces
us to assume that such particles (like the triplets) are at least as heavy
as the unification scale. It seems possible to arrange that in many models.

Based on the above arguments, we performed a complete scan over the
parameter space on which the gaugino masses depend -- \{$\delta $, $V_{7}$, $%
C_{2}$, $\alpha _{vis}^{-1}$\}, assuming
$Q-P=3,P_{\mathrm{eff}}=83$. As negative $\delta $ is necessary to
obtain the right unification scale, we take a
range $-10\leq \delta \leq 0$. $C_{2}$ is taken to be $\mathcal{O}(1)$. $%
\alpha _{vis}^{-1}\equiv \mathrm{Im}(f^{0})$ is taken to be of $\mathcal{O}%
(25)$. The lower and upper bounds on $V_{7}$ are given by \footnote{%
The upper bound is determined from the fact that in both heterotic and type
IIA duals of these vacua, the parameter $a\equiv \frac{V_{7}}{V_{Q}^{7/3}}$
is always less than unity \cite{Friedmann:2002ty}.}:
\begin{eqnarray}
V_{7}^{min} &=&1;\;\;\mathrm{weak\;supergravity\;constraint\;(Appendix\,\ref%
{constraints})}  \notag \\
V_{7}^{max} &=&V_{\hat{Q}_{vis}}^{7/3}\approx (\alpha _{unif}^{-1}-\delta
)^{7/3};\;\mathrm{corresponding\;to}\;a\equiv \frac{V_{7}}{V_{\hat{Q}%
_{vis}}^{7/3}}=1
\end{eqnarray}%
In addition, we consider a gravitino mass below $100~\mathrm{TeV}$ so that
the spectrum is light enough to be potentially be seen at the LHC, as well
as satisfy all our constraints. The analysis was done for two different
cases $\mu >0$ and $\mu <0$. As explained in section \ref{gauginospec}, $%
M_{1}$ and $M_{2}$ at low scale get contributions from the Higgs-Higgsino
loop (see eqn. (\ref{Higgs-Higgsino})) which depends on the sign of $\mu $.
This means that for a given $m_{3/2}$ which is consistent with all mass
bounds on the spectrum, $|M_{3}|/|M_{2}|$ for $\mu >0$ is greater than that
for $\mu <0$. So, it turns out that $\mu <0$ is excluded by gauge coupling
unification while $\mu >0$ is allowed. The gaugino mass spectra for $\mu >0$
compatible with precision gauge unification and all bounds on superpartner
masses are shown in Figure \ref{Fig:scangcu1}. The most stringent bound
among superpartner mass bounds is that of the lightest chargino from LEP II.
For the bino LSP case, the bound is $M_{\tilde{C}_{1}}\geq 104$ GeV. However
for a wino LSP, which turns out to be relevant for us, the bound depends on
the mass splitting $\Delta M\equiv M_{\tilde{\chi}_{1}^{\pm }}-M_{\tilde{\chi%
}_{1}^{0}}$; for simplicity we take $M_{\tilde{C}_{1}}\geq 80$ GeV.

The procedural details used are as follows. For a choice of $\delta$, $C_2$,
$V_7$ and $\alpha_{ unif}^{-1}$ in the above range as well as for a set of
initial values of Yukawa couplings and $z_{\mathrm{eff}}$ at $M_{\mathrm{unif%
}}\sim\mathcal{O}(10^{16})$ GeV, the MSSM spectrum was computed at low
scales using the analysis in sections \ref{gauginounif} and \ref{gauginospec}%
. The experimental values of the gauge couplings (both their max. and min.
values taking the uncertainty into account) were then RG evolved backwards
to the high scale using two-loop RGEs which depend on the superpartner
thresholds. The unified gauge coupling and the unification scale were
determined by the requirement $\alpha_1(M_{\mathrm{unif}})=\alpha_2(M_{%
\mathrm{unif}})$. The Yukawa couplings were evolved to the high scale at the
same time. The original parameters were scanned within their respective
ranges and only those values for which the initial assumed $\alpha_{\mathrm{%
unif}}$ was equal to the value of the computed $\alpha_{\mathrm{unif}}$ up
to experimental uncertainties, were recorded. $M_Z$ was checked to be
approximately $91~\mathrm{GeV}$. The condition for gauge coupling
unification, i.e. $\alpha_{3,min}(M_{\mathrm{unif}})< \alpha_{1,2}(M_{%
\mathrm{unif}})< \alpha_{3,max}(M_{\mathrm{unif}})$, was checked and only
sets of parameters which satisfied the above condition as well as other
constraints on superpartner masses, were recorded. In the above condition, $%
\alpha_{3,min}(M_{\mathrm{unif}})$ and $\alpha_{3,max}(M_{\mathrm{unif}})$
are the lower and upper values of $\alpha_3$ at the unification scale,
determined by RG evolving the low scale experimental value of $\alpha_3$
taking the uncertainties into account. The low scale gaugino mass spectra
consistent with precise gauge coupling unification are plotted in Figure \ref%
{Fig:scangcu1}. As explained earlier, this is only possible for $\mu > 0$.
One sees from Figure \ref{Fig:scangcu1} that only discrete values of gaugino
masses are possible since it is not possible to satisfy precision gauge
unification constraints for continuous sets of parameters.

\section{What is the LSP?}

\label{LSP}

From Figure \ref{Fig:scangcu1}, the lightest supersymmetric particle
(assuming $R$-parity conservation) turns out to be predominantly wino-like.
The Higgsinos are of $\mathcal{O}(m_{3/2})$ and are much heavier than the
gauginos. Here, as usual we have assumed that $Q-P=3$ and $P_{\mathrm{eff}%
}=83$.

It is worthwhile to compare and contrast the results obtained for the $G_{2}$%
-MSSM for the nature of the LSP with those for the Type IIB vacua
corresponding to the \textquotedblleft mirage mediation" framework
mentioned in the introduction. There one always gets bino LSPs. In
mirage mediation, the gaugino mass contribution is dominated by
the tree-level and conformal anomaly contribution (the first term
in (\ref{anom1})) which are of the same order \cite{Choi:2007ka}.
The second and third term in (\ref{anom1}) are negligible because
of the assumption of sequestering. On the other hand, for the
$G_{2}$-MSSM, the Konishi anomaly contribution coming from the
second and third term in (\ref{anom1}) is also important as one
does not expect sequestering in general.
The
second important difference is that the $\mu $ parameter is very
large for the $G_{2}$-MSSM (of $\mathcal{O}(10)$ TeV) compared to
that for mirage mediation. This implies that the finite
contribution to $M_{1}$ and $M_{2}$ from (\ref{Higgs-Higgsino}) in
the $G_{2}$-MSSM is quite important in contrast to that in mirage
mediation. Also, as seen from the bottom plot in Figure
(\ref{Fig:gaugino-low1}), $\mu <0$ gives rise to bino LSPs;
however
only $G_{2}$-MSSM vacua with $\mu >0$ should be considered since those with $%
\mu <0$ are disfavored by precision gauge coupling unification as explained
in section \ref{gaugino-gcu}. Therefore, due to all the above reasons, the
nature of the LSP obtained for the $G_{2}$-MSSM is different from that for
mirage mediation.

\subsection{Dark Matter Relic Abundance and Cosmological Evolution of Moduli}

It is well known that wino LSPs annihilate quite efficiently
compared to bino LSPs due to the larger SU(2) gauge coupling
$g_2$. Therefore, for wino masses of $\mathcal{O}$(100-500) GeV,
as is natural for the $G_2$-MSSM vacua, the thermal contribution
to the relic density of these wino LSPs is much smaller ($\sim
0.01-0.1$ times) than the observed upper bound on the relic
density. This implies that if (wino) LSPs constitute most of the
Dark-Matter(DM) in the universe, they must be produced
non-thermally. The issue of non-thermal production mechanisms of
these LSPs is intricately linked to the cosmological evolution of
moduli after inflation. As it turns out, during the course of
their evolution moduli decay to LSPs giving rise to an appreciable
non-thermal contribution which could give rise to a wino LSP relic
density of approximately the right amount. This will be reported
in detail in a future study \cite{moduligravitino}. Here we will
just outline the salient features of our analysis of cosmological
moduli evolution.

It is well known that light gauge-singlet scalar fields such as moduli and
meson fields couple very weakly (only gravitationally) to the visible sector
causing them to decay at late times which in turn could spoil the successful
predictions of Big-Bang Nucleosynthesis (BBN). This is a generic problem in
string/$M$ theory compactifications and has to be addressed carefully. In $%
G_{2}$ compactifications, since all moduli are stabilized, the moduli mass
matrix, as well as all couplings of the moduli to the visible sector can be
explicitly computed in terms of the microscopic parameters. It turns out
that there is a hierarchy of mass scales for the moduli with one modulus
being much heavier than the others, the lighter ones being $\mathcal{O}%
(2m_{3/2})$. To set the stage for the analysis, it is reasonable to assume
that enough inflation occurs at the end of which reheating gives rise to a
radiation dominated universe with the moduli displaced from their minimum
values. Then, one has to look at the evolution of these moduli carefully,
taking into account the hierarchy of scales involved. During the course of
their evolution, the moduli will start coherent oscillations and then decay
to visible particles ultimately leading to SM particles and LSPs. Since all
the relevant moduli-matter couplings can be explicitly computed,
this sequence of steps
can be carried out reliably and the LSP relic density can be computed. It
turns out that the wino LSP relic density for natural values of parameters
is in the range 0.1-1.
The cosmological evolution of moduli in
$G_{2}$-MSSM vacua is, in a sense, an explicit realization of the
Randall-Moroi scenario \cite{Moroi:1999zb} in which the moduli are
quite heavy ($10-100$ TeV) and decay before BBN to wino-like LSPs.
The detailed spectrum is however different and more detailed
computations can be done for the $G_{2}$-MSSM vacua.

\section{Benchmark Spectra and Characteristic
Features}\label{features}

\begin{table}[t!]
\label{tab:MG2}
\begin{tabular}{||c||c|c|c|c||}
\hline\hline parameter\rule{0pt}{3.0ex}\rule[-1.5ex]{0pt}{0pt} &
Point~1 & Point~2 & Point~3 & Point~4 \\ \hline\hline
$\delta$ \rule{0pt}{3.0ex} & $\qquad$ -4 $\qquad$ & $\qquad$ -6 $\qquad$  & $%
\qquad$ -8 $\qquad$ & $%
\qquad$ -10 $\qquad$\\
$m_{3/2} $ \rule{0pt}{3.0ex} & 67558 & 35252 & 34295 &
17091\\
\hline
$V_7$ \rule{0pt}{3.0ex} & 14 & 21.6  & 22 & 35\\
$\alpha_{unif}^{-1}$ \rule{0pt}{3.0ex} & 26.7 & 26.4 & 26.5 & 26.0\\
$Z_{\mathrm{eff}}$ \rule{0pt}{3.0ex}  & 1.58 & 1.65 & 1.65 & 1.77\\
\hline
$\tan\beta$ & 1.44 & 1.45 & 1.45 & 1.45\\
$\mu $ & 87013 & 45572 & 44164 & 22309\\
$m_{\tilde g}$ & 994.7 & 732.5 & 900.4 & 573.5 \\
$m_{\widetilde \chi_1^0}$ & 116.6 & 110.9 & 173.1 & 107.1 \\
$m_{\widetilde \chi_2^0}$ & 390.0 & 228.3 & 253.5 & 137.1\\
$m_{\widetilde \chi_1^{\pm}}$ & 116.7 & 111.0 & 173.2 & 107.3\\
$m_{\tilde u_L}$ & 67600 & 35254 & 34298 & 17094\\
$m_{\tilde u_R}$ & 67559 & 35264 & 34298 & 17093\\
$m_{\tilde t_1}$ & 18848 & 9010  & 8700 & 3850\\
$m_{\tilde t_2}$ & 49554 & 25707 & 24998 & 12378\\
$m_{\tilde b_1}$ & 49554 & 25707 & 24998 & 12378\\
$m_{\tilde b_2}$ & 67497 & 35220 & 34265 & 17076\\
$m_{\tilde e_L}$ & 67558 & 35253 & 34296 & 17091\\
$m_{\tilde e_R}$ & 67559 & 35253 & 34296 & 17091\\
$m_{\tilde \tau_1}$ & 67527 & 35237 & 34280 & 17084\\
$m_{\tilde \tau_2}$ & 67543 & 35245 & 34288 & 17088\\
$m_h$ & 123.6 & 120.8 & 120.3 & 118.1 \\
$m_A$ & 134083 & 70053 & 68031 & 34107\\
$A_t$ & 14267 & 6208 & 6024 & 2379\\
$A_b$ & 3114 & 1637 & 1604 & 805\\
$A_{\tau}$ & 1935 & 972 & 965.5 & 468.7\\ \hline\hline
\end{tabular}%
\caption{``Microscopic" parameters and low scale spectra for four benchmark $%
G_2$-MSSM models. All masses are in GeV. The top mass is taken to
be $174.3$ GeV in our calculation. For all the above points,
$Q-P=3$ and $\P=83$ are taken as discussed in the text. The
gravitino mass depends mainly on the combination $C_2 V_7^{-3/2}$
as in Eq. (\ref{grav-approx}). So the spectra are largely
determined by two parameters $\delta$ and $m_{3/2}$. All the above
spectra are consistent with current observations. Scalar masses
are lighter for benchmark 4, so flavor changing effects need to be
explicitly checked later.}
\end{table}

\begin{figure}[h!]
\begin{center}
\epsfig{file=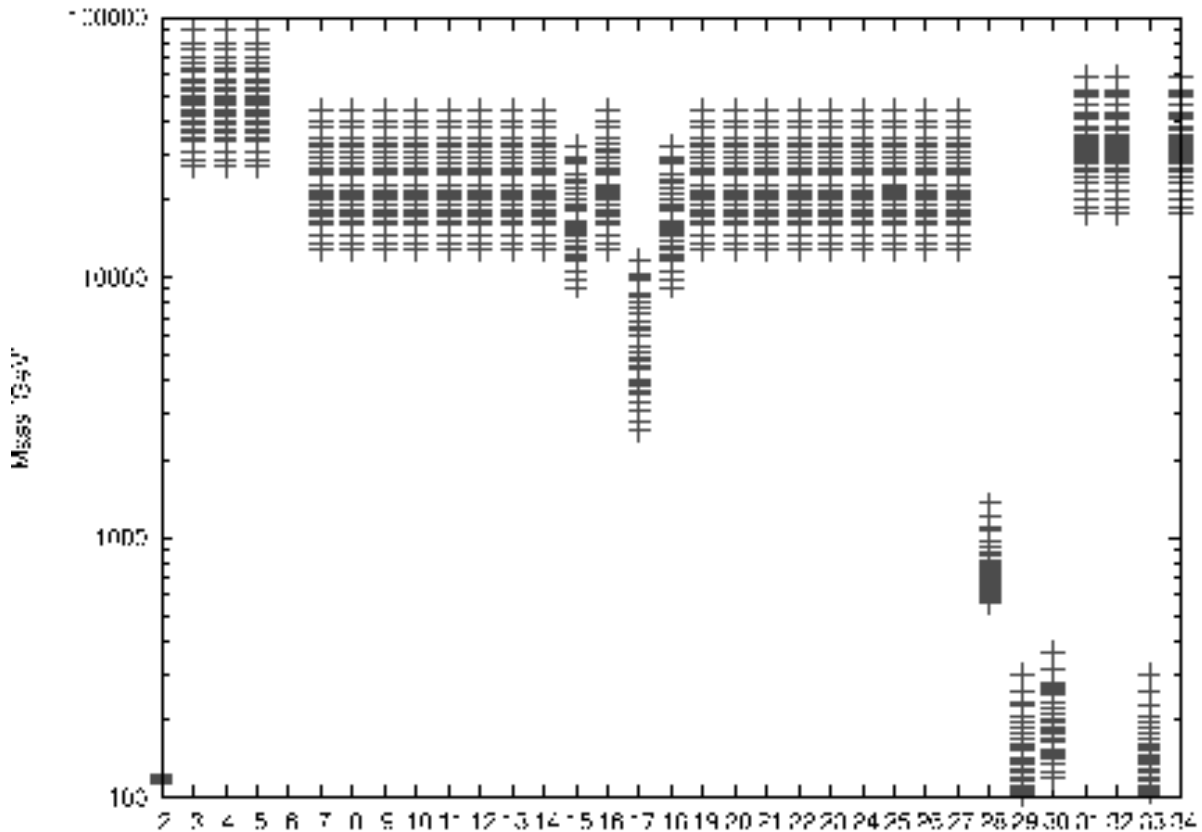,height=11cm, angle=0}
\end{center}
\caption{Semi-log Plot showing variations in the spectra of $G_2$-MSSM
models. The columns correspond to the following masses. 2: Light Higgs (h).
3-5: Heavy Higgs/pseudo-scalar Higgs. 7-14: 1st/2nd Generation squarks.
15-18: 3rd Generation squarks. 19-27: Sleptons. 28: gluino. 29: $\tilde{N}_1$%
. 30: $\tilde{N}_2$. 31-32: $\tilde{N}_{3-4}$ . 33-34: $\tilde{C}_{1-2}$.}
\label{fig:spectrum}
\end{figure}

In this section, we take a brief look at the pattern of spectra
obtained in the $G_2$-MSSM and point out their main features. A
detailed study of the collider phenomenology of the $G_2$-MSSM
will be reported in \cite{LHC-G2}. Table I shows the `microscopic'
input parameters and corresponding low-scale spectra for four
benchmark $G_2$-MSSM models while Figure \ref{fig:spectrum}
displays the physical masses for many $G_2$-MSSM models. As
discussed in section \ref{spectra}, the $G_2$-MSSM spectrum is
characterized by heavy multi-TeV scalars and Higgsinos, sub-TeV
gauginos,
and an SM-like Higgs field\footnote{%
The parameter $\tan(\beta)$ is of $\mathcal{O}(1)$.}. Thus, the arrangement
of the sub-TeV fields crucially determines the pattern of observable
signatures at the LHC.

$G_2$-MSSM models with $Q-P=3,P_{\mathrm{eff}}=83$ and which are consistent
with precision gauge coupling unification give rise to wino LSPs. For this
case, the following hierarchy between the sub-TeV particles is observed:
\begin{eqnarray}
m_{\tilde{g}} > m_{\tilde{N_2}} > m_{\tilde{C_1}} > m_{\tilde{N_1}}
\end{eqnarray}
The $\tilde{N_1}$ is nearly degenerate with the $\tilde{C}_1$ ($m_{\tilde{C_1%
}} - m_{\tilde{N_1}} \lesssim 200$ MeV). Variations of high-scale input
parameters $V_7$ and $\delta$ simply shift the overall mass scale of the
fields, but do not modify this hierarchy. This feature significantly
constrains the possible decay modes observable at the LHC. For example, the
dominant production modes are $\tilde{C}_1\tilde{C}_1$ and $\tilde{C}_1%
\tilde{N}_1$ followed by $\tilde{g}\tilde{g}$. The decay of the charginos to
the LSP is characterized by soft pions. The charginos tend to decay inside
the detector giving rise to short charged track stubs. A systematic study of
these signatures in both ATLAS and CMS is required to properly estimate the
discovery potential of these decays.
The gluinos on the other
hand decay dominantly via a three-body decay to $t\bar{t}\tilde{N}_2$ since
the lightest stop is mostly right handed. This mode can give rise to
signatures with many b-jets from multi-top production. Other decay modes,
although suppressed, are also available. A study of these issues is underway
and will be reported in \cite{LHC-G2}.

The tree level production rate for $\tilde{C}_{1}$ +
$\tilde{N}_{2}$ would vanish for pure bino $\tilde{N}_{2}$ so the
size of this cross section is a measure of the wino part of the
LSP. \ Similarly, the rate for production of $\tilde{C}_{1}$  +
$\tilde{N}_{1}$ is about two times larger for a wino LSP than for
a higgsino LSP and can thus help fix the LSP type. \ The signal
for us is the single soft charged pion, which will not trigger,
but can be triggered in association with an initial state photon
or jet, and the chargino will pass a few layers of the vertex
detectors so in principle these rates might be accessible.

Since all the Higgs bosons except the light one ($h$) will be in
the TeV range, and undetectable, it is interesting to have ways to
distinguish the light Higgs ($h$) from the SM one ($h_{SM})$. One
way to do that in principle is to exploit the chargino loops in
the modes $h \rightarrow \gamma \gamma $ and $h \rightarrow
Z\gamma ,$ which will make the branching ratio different from the
SM one. \ Accurate measurements would be needed, and effects of CP
violation would have to be untangled to obtain definitive results.

Finally, we mention that the fit to EW precision observables
apparently  improves with light charginos and neutralinos
\cite{Martin:2004id}.

\section{Conclusions and Future Directions}

\label{conclude}

In this paper, we have studied a framework arising from the
low-energy limit of $M$ theory which gives rise to vacua in which
the moduli can be stabilized and a stable
hierarchy between the electroweak and Planck scales can be
generated. A well-motivated phenomenological model - the
$G_{2}$-MSSM, can be naturally defined within this framework and
its properties can be studied in detail. The model arises by
compactifying $M$ theory on a seven dimensional singular manifold
of $G_{2}$ holonomy. Strong gauge dynamics in the hidden sector
simultaneously stabilizes all the moduli and breaks supersymmetry.
With matter fields in at least one of the hidden sectors, there is
a de Sitter minimum that is unique for a given choice of
$G_2$-manifold. We look for
solutions which are within the supergravity regime, where the
volume of the three-manifold supporting the hidden sector is
large, the number of moduli is not constrained to be small, and
the cosmological constant can be tuned to zero.

Then, remarkably, we find that all solutions have $m_{3/2}$ less
than a few hundred TeV and the suppression of gaugino masses leads
to TeV and sub TeV scale new physics.
The
requirement that the cosmological constant can be tuned to zero is
non-trivial. This constraint  strongly depends on the
nature of the three-manifolds on which the hidden sectors are
supported. At present this computation has been carried out
only for a very special class of three-manifolds
${\bf S^3}/{\bf Z_k};\,k\in {\bf Z}$,
where it turns out that the above requirement can only be
satisfied for large values of $k$, which may not be very natural.
However, as discussed in Appendix \ref{Peff}, various other
possibilities for three-manifolds exist which might help in
satisfying the requirement more naturally.

Further, we find that tree level
gaugino masses are suppressed compared to $m_{3/2}$ by a large
factor, approximately 83. This occurs mainly because the largest
supersymmetry breaking $F$-term is that from hidden sector meson
fields, which do not contribute to the gaugino masses. No such
suppression occurs generically for scalar masses or trilinears,
which are therefore of order $m_{3/2}$.
Since the Kahler potential is not sequestered, the scalar masses are
expected to be of order $m_{3/2}$, with the gaugino masses much
smaller. There is no theoretical lower limit on $m_{3/2}$, but the
absence of charginos at LEP gives a lower limit on the gaugino
masses $M_{1}$ and $M_{2}$, and therefore a lower limit on
$m_{3/2}$, which is of order 10-50 TeV. So, scalars are predicted
to be that heavy.

We focus on solutions within the $G_2$-MSSM which are consistent
with precision gauge coupling unification. Then the LSP is
wino-like. The $G_2$-MSSM framework is broadly similar to other
models with heavy scalars and light gauginos with wino LSPs
\cite{Asai:2007sw}, but there are differences in details. As is
well known, the thermal relic density of wino LSPs is small
compared to the observed one. However, many LSPs are generated
from moduli decay, along with significant entropy. When the
cosmological evolution of moduli is taken into account the
resulting relic density is near the observed one. The moduli
masses and lifetimes can be computed in detail in terms of the
microscopic details, and we find a nice explicit realization of
the Moroi-Randall \cite{Moroi:1999zb} mechanism\footnote{In
details, the $G_2$-MSSM differs from the Moroi-Randall scenario,
however.}. The predicted LHC signatures are characteristic and
interesting. RG evolution down to the gluino mass $\lesssim $ a
TeV leads to the lightest squark being a mainly right handed stop,
which itself decays mainly to the top and the second neutralino,
which then decays to the $W$ and the lightest chargino (wino).
Thus, for pair produced gluinos, there are many events
corresponding to a final state with four tops, large missing
energy, and two charginos, dramatic signatures that are easily
triggered on and distinguishable from background. We also initiated
analysis of the CP violating phases of the theory, which are
surprisingly simple.

There are many possibilities for future research. From a
theoretical perspective, one of the most outstanding problems is
to construct global examples of $G_{2}$ manifolds with the right
structure of conical and orbifold singularities. This would
require a major breakthrough from a mathematical point of view.
However, a better understanding of the dualities from the
heterotic and Type IIA string could also lead to important
insights. Another important theoretical issue is to better
understand the assumptions made about the K\"{a}hler potentials
and study possible generalizations. Although we have checked that
our main qualitative results about the suppression of gaugino
masses do not depend on the detailed form of the K\"{a}hler
potential $K$, further insights would be welcome.

From a phenomenological perspective, it would be extremely useful
to study variants of the minimal proposal which could solve
important phenomenological problems while still retaining the
desirable features. A good feature of this framework is that
important phenomenological questions such as inflation, generation
of neutrino masses, explanation of Yukawa couplings and the origin
of flavor\footnote{This is intrinsically related to the issue of
the K\"{a}hler potential mentioned above.}, the matter asymmetry,
the strong CP problem, the little hierarchy problem, etc. can all
be addressed within this framework.

\acknowledgments The authors appreciate helpful conversations with
and suggestions from Jacob Bourjaily, Kiwoon Choi, Joseph Conlon,
David Morrissey, Brent Nelson, Aaron Pierce, Liantao Wang and
James Wells. The research of KB, GLK, PK and JS is supported in
part by the US Department of Energy. GLK appreciates support and
hospitality from the Institute for Advanced Study including a
grant from the Ambrose Monell Foundation. PK and BA thank Rutgers
University for their hospitality. JS acknowledges support from
Princeton University and hospitality from the Institute of
Advanced Study.

\appendix

\section{Computation of $P_{\mathrm{eff}}$}

\label{Peff}

As we have seen in section \ref{summary}, a large $P_{\mathrm{eff}}$ is
crucial for the validity of our solutions and the supergravity
approximation. It also leads to the the suppression of tree-level gaugino
masses compared to the gravitino mass. Finally, in order to tune the
cosmological constant, one requires $P_{\mathrm{eff}} = 84$ (83 if one
includes higher order corrections) for $Q-P=3$. In this paper, until now we
have just used $P_{\mathrm{eff}}$ as one of the parameters which could vary
within a certain range. However, in an explicit microscopic construction of
the hidden sector, it is computable from first principles. From the
definition:
\begin{eqnarray}
P_{\mathrm{eff}}\equiv P\ln\left(\frac{C_1}{C_2}\right)  \label{Peff-def}
\end{eqnarray}
it is easy to see that for a large $P_{\mathrm{eff}}$ such as 83,
a large splitting in the coefficients $C_1$ and $C_2$ is required.
At tree level, these coefficients are simply determined by the
cutoff scale of the effective gauge theory and are given by
$(\Lambda_{\mathrm{cutoff}}/M_P)^3$. This would not give a large
$P_{\mathrm{eff}}$. However, one has to take into account
threshold corrections to the gauge couplings of the hidden
sectors. To compute the threshold corrections, one has to specify
the concrete setup of the hidden sector $\hat{Q}$, as well as the
geometry of the three-manifold where the hidden sector lives.
Generally the one-loop gauge couplings can be written as:
\begin{eqnarray}
\frac{16\pi^2}{g^2({\mu})}=\frac{16\pi^2}{g_M^2}+b\log(\Lambda^2/\mu^2)+S,
\end{eqnarray}
where $b$ is the one-loop beta function coefficient, and $S$ is the one-loop
threshold correction. For instance, the contribution from KK modes has the
form \cite{Friedmann:2002ty}:
\begin{eqnarray}
S=S^{\prime }+2N_c \log({\rm Vol}(\hat{Q})
\Lambda_{\mathrm{cutoff}}^3),
\end{eqnarray}
where ${\rm Vol}(\hat{Q})$ is the volume of the hidden-sector
three-manifold $\hat{Q}$ and $S^{\prime }$ can be expressed in
terms of certain topological invariants of $\hat{Q}$, known as the
``Ray-Singer analytic torsion". Before we go to explicit examples,
we would like to show the general requirement on the threshold
corrections. Let us first denote
the gauge kinetic function as $f=f_{0}+\Delta f_1$ and $f_2=f_{0}+\Delta f_2$%
, where $\Delta f_{1,2}$ are the corresponding threshold
correction. The superpotential from strong gauge dynamics can be
written as:
\begin{eqnarray}
W \sim \Lambda_{\mathrm{cutoff}}^{3+\alpha}\, |\phi|^{-\alpha}\,P\, e^{-\frac{2\pi}{P}%
(f+\Delta f_1)}+\Lambda_{\mathrm{cutoff}}^3\, Q\,e^{-\frac{2\pi}{Q}(f+\Delta
f_2)}
\end{eqnarray}
We can easily identify the coefficient $C_{1,2}$ as follows:
\begin{eqnarray}
C_1&=& \left(\frac{\Lambda_{\mathrm{cutoff}}}{M_P}\right)^{3+\alpha} e^{-\frac{%
2\pi}{P}\Delta f_1} \\
C_2&=&\left(\frac{\Lambda_{\mathrm{cutoff}}}{M_P}\right)^{3} e^{-\frac{2\pi}{%
P}\Delta f_2}
\end{eqnarray}
Since $\alpha=2/P$ is small, we have
\begin{eqnarray}
\frac{C_1}{C_2}\approx e^{-\frac{2\pi}{P}\Delta f_1+\frac{2\pi}{Q}\Delta
f_2}.  \label{rat-C1-C2}
\end{eqnarray}
For the case $Q-P=3$ and $P_{\mathrm{eff}}=84$, using Eq.(\ref{Peff-def})
and (\ref{rat-C1-C2}) we have the estimate
\begin{eqnarray}
\Delta f_1 -\Delta f_2\sim 14.  \label{thr-dif}
\end{eqnarray}
In view of the fact that $f_0\approx \frac{14}{3}Q=\mathcal{O}(50)$, the
requirement Eq.(\ref{thr-dif}) is not completely unreasonable.

\subsection{A Particular Example - $\hat{Q}=S^3/Z_k$}

As a particular example, we consider the three-manifold $\hat{Q}$ to be the
lens space $S^3/Z_k$ as in this case the threshold corrections can be
computed. In addition, for concreteness, we consider a situation where the
first hidden sector gauge group is obtained from a larger group $SU(P+M+1)$
by Wilson line breaking $SU(P+M+1)\rightarrow SU(P+1)\times SU(M)\times U(1)$%
, while the second hidden sector group is still $SU(Q)$ without breaking.
Again we assume one flavor of charged matter $Q$ and $\bar Q$. As long as $M$
is sufficiently smaller than $P$ (such as $< P/2$), we can neglect its
contribution to the superpotential and also in moduli stabilization. The
calculation of the threshold correction is similar to that in \cite%
{Friedmann:2002ty}. For the first hidden sector, it is given by $S_1^{\prime
}= 2(P+1) \,T_{\mathcal{O}}+2M\, T_{\lambda}$, while for the second one it
is $S_2^{\prime }=2Q\, T_{\mathcal{O}}$. $T_{\lambda}$ and $T_{\mathcal{O}}$
are the relevant torsions:
\begin{eqnarray}
T_{\mathcal{O}}=-\log(k), \quad T_{\lambda}=\log(4\sin^2(G\pi \lambda/k)),
\label{torsion}
\end{eqnarray}
where $G=P+M+1$, and $\lambda$ is an integer specifying the Wilson line. As
discussed above, $C_{1,2}$ can be calculated straightforwardly, which are
\begin{eqnarray}
C_1&=& M_P^{-3}\,\langle\mathrm{Vol}(\hat{Q})^{-(1+1/P)}\rangle\,
\Lambda_{\mathrm{cutoff}}^{-1/P}\, e^{-\frac{S_1^{\prime }}{2P}} \\
C_2&=& M_P^{-3}\,\langle\mathrm{Vol}(\hat{Q})^{-1}\rangle\,
e^{-\frac{S_2^{\prime }}{2Q}}.
\end{eqnarray}
Here, ${\rm Vol}(\hat{Q})$ is the dimensionful volume of the
hidden-sector three-manifold $\hat{Q}$. It is important to
remember that $C_1,C_2$ should be thought of as depending on the
vacuum expectation value of ${\rm Vol}(\hat{Q})$ \footnote{i.e.
obtained after moduli stabilization} as shown in the above
equation. Therefore, this dependence does not invalidate the
holomorphicity of the superpotential. One can now compute the
$P_{\mathrm{eff}}$ from the above equations:
\begin{eqnarray}
P_{\mathrm{eff}} &\approx&P (-\frac{S_1^{\prime }}{2P}+\frac{S_2^{\prime }}{%
2Q})  \notag \\
&=&-\,T_{\mathcal{O}}-M\, T_{\lambda}  \notag \\
&=&\log(k)-M\log(4\sin^2(G\pi \lambda/k))
\end{eqnarray}
It is obvious that $k$ has to be very large to get a large
$P_{\mathrm{eff}}$. For example, $P=15$, $Q=18$, $M=10$, $V_7=50$,
$\lambda=80$ and $k=99$ gives $P_{\mathrm{eff}}=58$ and
$C_2=1.5\times 10^{-4}$. Notice $C_2$ is much smaller than one.
The gravitino for this case is about $0.8 \mathrm{TeV}$. One can
also consider other patterns of symmetry breaking, e.g. $SO(2(P+1))%
\rightarrow SU(P+1)\times U(1)$. In this case smaller values of
$k$ compared to the previous example can give rise to a large
$\P$, although in general a large $k$ is still needed. Large
values of $k$ may seem unrealistic, but that is not clear since
the allowed possibilities for compact $G_2$ manifolds fibred over
Lens spaces with large $k$ are not known. In addition, although at
present it is not known how to compute the torsion for other
three-manifolds, it is possible that a large $P_{\mathrm{eff}}$
can be obtained more ``naturally'' in other examples.

\subsection{More General Possibilities}

Other three manifolds might give rise to a large $P_{\mathrm{eff}}$ more naturally.
Rather than study further explicit examples we give a toy model which illustrates
this possibility. The model has parameters which extend the previous example and
is given by
\begin{eqnarray}  \label{general}
T_{\mathcal{O}}= -\gamma_0\log(k), \quad
T_{\lambda_1}=\gamma_1\log(\alpha\sin^2(G\pi \lambda_1/k)),\;
T_{\lambda_2},\; T_{\lambda_3},...
\end{eqnarray}
where $G$ is an integer and
$%
\gamma_{0,1}$ are determined by the topology of the manifold and are kept as
free parameters. $\alpha$ is determined by group theory. In general, there
could be other non-trivial torsions $T_{\lambda_2},T_{\lambda_3}$, etc.
depending on how the higher gauge group is broken by the Wilson lines. In
order to illustrate the idea, we will restrict to $T_{\mathcal{O}}$ and $%
T_{\lambda_1}$. Let's again consider the case where the first hidden sector
group $SU(P+1)$ arises from the breaking $SU(P+M+1)\rightarrow SU(P+1)\times
SU(M)\times U(1)$. Now, it is possible to get both $P_{\mathrm{eff}}\approx
84$ and $C_2\sim \mathcal{O}(1)$. For example, the set of parameters $%
\gamma_0=\gamma_1=6.3$, $P=15$, $Q=18$, $M=10$, $V_7=50$ and $k=11$ gives $%
P_{\mathrm{eff}}=84.2$ and $C_2=5.4$.

The above example was shown just to illustrate the fact that with more
general three-manifolds $\hat{Q}$, it may be possible to obtain a large $P_{%
\mathrm{eff}}$ quite naturally. As another possibility, if $\hat{Q}$ is such
that the relevant torsions $T_{\mathcal{O}},T_{\lambda_1},T_{\lambda_2}$,
etc. in (\ref{general}) have a linear dependence on $k$ instead of
logarithmic one, it is quite easy to obtain a large $P_{\mathrm{eff}}$
naturally. In addition, if there are massive quarks\footnote{%
Of course, they should be heavier than the strong coupling scale}
which are charged under the hidden sector gauge group, the strong
coupling scale will be lowered and both the values of
$P_{\mathrm{eff}}$ and $C_2$ will be affected.

To summarize, the values of $P_{\mathrm{eff}}$ and $C_2$ depend
crucially on the microscopic details of the hidden sector and can
take a wide range of values. Therefore, in our phenomenological
analysis, we have simply assumed that $P_{\mathrm{eff}}$ and $C_2$
can take values in the range of phenomenological interest.

\section{Constraints on ``microscopic" parameters}

\label{constraints}

The vacua in realistic $G_2$ compactifications are characterized by the
``microscopic" parameters $\{N_i$, $N_i^{sm}$, $a_i$, $N$, $P$, $Q$, $C_1$, $%
C_2$, $\delta;\,i=1,2,..,N\}$. Phenomenologically relevant quantities
however, are sensitive to some parameters directly such as $P$, $Q$, $C_2$
and $\delta$ and some of them in combinations such as $V_7$, $V_{\hat{Q}%
_{vis}}$ and $P_{\mathrm{eff}}$. As mentioned in section \ref{soft-unif},
these combinations have to satisfy various constraints such as the
``supergravity constraint", the ``dS vacuum constraint", the ``unified
coupling constraint" and the ``unification scale constraint". As promised,
here we discuss the first three in detail. The unification scale constraint
has already been discussed in section \ref{gaugino-gcu}.

We would like to find sets of microscopic parameters which give rise to $%
\{V_7,V_{\hat{Q}_{vis}},P_{\mathrm{eff}}\}$ such that the above mentioned
constraints can be satisfied. In order to take into account the effects of
the parameters $a_i,N_i,N_i^{sm},i=1,..,N$, we consider the following two
extreme cases:
\begin{eqnarray}
&&\text{(a) All $a_i$ are roughly equal, so $a_i\approx \frac{7}{3N}$.}
\notag \\
&&\text{(b) A few $a_i$ are much larger than the rest, for simplicity we
take $a_1\approx \frac{7}{3}$ and $a_{i\neq 1}\approx0$.}  \notag
\end{eqnarray}
Since other choices of the above parameters lie in between the two extremes,
presumably so would their implications. For later use, it is useful to note
that at the meta-stable dS minimum, the moduli are stabilized at \cite%
{Acharya:2007rc}:
\begin{eqnarray}
s_i&=&\frac{a_i}{N_i}\nu,\quad \nu\approx\frac{3P_{\mathrm{eff}} Q}{14\pi
(Q-P)}
\end{eqnarray}
where $P_{\mathrm{eff}} \equiv P\,\log\left(\frac{C_1}{C_2}\right)$. Let us
first consider the weak supergravity constraint for the phenomenologically
interesting dS vacua, which reads:%
\begin{eqnarray}
V_7&=&\prod_{i=1}^{N}s_i^{a_i} > 1
\end{eqnarray}
For case (a), we can rewrite the seven dimensional volume as
\begin{eqnarray}
V_7&\approx&\left(\frac{7\nu}{3N}\right)^{7/3}\left(\prod_{i=1}^{N}
N_i\right)^{-\frac{7}{3N}}  \notag \\
&\approx&\left(\frac{7\nu}{3N{\bar N}}\right)^{7/3}
\end{eqnarray}
where ${\bar N}$ is defined to be the geometric mean of the $N_i$'s. One
should keep in mind that $\bar N$ can be small even if some of the $N_i$'s
are large, if $N$ is $\mathcal{O}(10)$ or greater. The supergravity
constraint then turns out to be:
\begin{eqnarray}
\frac{7\nu}{3N{\bar N}}> 1 \quad \Longrightarrow \quad \frac{ {P_{\mathrm{eff%
}}} Q}{2\pi N \bar{N} (Q-P)} > 1  \label{Eq:sugra-constraint1}
\end{eqnarray}
For the case (b), we have $\left(\frac{a_i}{N_i}\right)^{a_i} \approx 1$ for
$i\neq 1$, and so $V_7=\left(\frac{a_1\nu}{N_1}\right)^{7/3}$. Therefore the
constraint turns out to be
\begin{eqnarray}
\frac{a_1}{N_1}\nu > 1 \quad \Longrightarrow \quad \frac{{P_{\mathrm{eff}}} Q%
}{2\pi N_1 (Q-P)} > 1  \label{Eq:sugra-constraint2}
\end{eqnarray}
A typical set of ``reasonable'' as well as phenomenologically interesting
values is $P_{\mathrm{eff}}\sim \mathcal{O}(10-100)$, $Q\sim \mathcal{O}(10)$%
, $Q-P\sim \mathcal{O}(1)(\mathrm{but}\,> 3)$, $\bar{N} \sim \mathcal{O}(1)$
and $N_1\sim \mathcal{O}(1)$, which easily satisfies the supergravity
constraint in case (b). The supergravity constraint for case (a) is also
satisfied for many sets of values of the parameters in the above ranges,
although not as easily for case (b). For a general case which lies between
(a) and (b), we expect a situation in between the two and the constraint
should be satisfied for parameters in the above range. The important point
is that this constraint can always be satisfied by an asymmetric
distribution of $a_i$.

Now let us consider the unified coupling constraint. The gauge kinetic
function for the visible sector is the volume of the visible three-cycle
\begin{eqnarray}
V_{\hat{Q}_{vis}}= \sum_{i=1}^{N} N_i^{vis} s_i,  \label{Eq:ineq2}
\end{eqnarray}
which obeys the following inequality:
\begin{eqnarray}
V_{\hat{Q}_{vis}} > N \left(\prod_{i=1}^{N} N_i^{sm}\right)^{1/N}
\left(\prod_{i=1}^{N} s_i\right)^{1/N}  \label{Eq:vol-constrant1}
\end{eqnarray}
For case (a), (\ref{Eq:ineq2}) can be written as:
\begin{eqnarray}
V_{\hat{Q}_{vis}} > N \left(\prod_{i=1}^{N} N_i^{sm}\right)^{1/N}\left(\frac{%
7\nu}{3N{\bar N}}\right)
\end{eqnarray}
From Eq.(\ref{Eq:sugra-constraint1}) and assuming $N_i^{sm} > 1$ for all $i$%
, we find $V_{\hat{Q}_{vis}} > N$. Since for the MSSM $V_{\hat{Q}%
_{vis}}=\alpha_M^{-1}\sim \mathcal{O}(25)$, this implies $N\lesssim \mathcal{%
O}(25)$. Thus, equal values of $a_i$ require the number of moduli to be
relatively small. One way out of this is that most of the $N_i^{sm}$ are
zero, and only $p\lesssim \mathcal{O}(10)$ of them are nonzero. This
however, is non-generic. For case (b), one has $\left(\prod_{i=1}^{N}
s_i\right)^{1/N}\sim 0$. Therefore, inequality (\ref{Eq:vol-constrant1}) can
be easily satisfied. Again, case (b) is more easily satisfied than case (a).
For a more general situation lying in between (a) and (b), one expects that
the constraint can be satisfied for many sets of values of the microscopic
parameters.

Finally, the dS vacuum constraint sets an upper limit on $P_{\mathrm{eff}}$:
\begin{eqnarray}
{P_{\mathrm{eff}}}<\frac{28(Q-P)}{3(Q-P)-8}
\end{eqnarray}
There is also a rough lower limit on $P_{\mathrm{eff}}$. From the
supergravity constraint for cases (a) and (b), one gets:
\begin{eqnarray}
P_{\mathrm{eff}} > \frac{2\pi N\bar{N}(Q-P)}{Q}\;\;\;\mathrm{case\;(a)}
\notag \\
P_{\mathrm{eff}} > \frac{2\pi N_1(Q-P)}{Q}\;\;\;\mathrm{case\;(b)}
\end{eqnarray}
For $Q-P=3$ \footnote{%
This is preferred phenomenologically as will be seen soon.}, $Q=\mathcal{O}%
(10)$, $N=\mathcal{O}(50)$ and $\bar{N}=\mathcal{O}(1)$, $P_{\mathrm{eff}}
\gtrsim 50$ for case (a) and $P_{\mathrm{eff}} \gtrsim 1$ for case (b). So
we can see the lower limit from the supergravity condition is really mild.
However for our solution to be valid, we should always restrict to large $P_{%
\mathrm{eff}}$, which should be $\sim 50$. For fixed $Q-P$ and $V_7$, the
lightest gravitino mass is achieved when ${P_{\mathrm{eff}}}$ reaches its
maximum which happens when the cosmological constant is tuned to be zero. By
increasing $Q-P$, the upper limit of ${P_{\mathrm{eff}}}$ decreases, e.g.,
\begin{eqnarray}
Q-P&=&3,\quad {{P_{\mathrm{eff}}}}<84,  \notag \\
Q-P&=&4,\quad {{P_{\mathrm{eff}}}}<28,  \notag \\
Q-P&=&5,\quad {{P_{\mathrm{eff}}}}<20.
\end{eqnarray}
Therefore solutions with $Q-P =3$ and $\P =83$ are required in
order to have a generic number of moduli and a TeV scale gravitino
mass, and to tune the cosmological constant.

\section{Threshold corrections to Gaugino masses from Heavy Fields}

\label{thresholdcorr}

In the following, we show explicitly that the presence of a heavy fields
will not change the gaugino masses at low energy. The relevant terms in the
Lagrangian at $M_{\mathrm{unif}}$ can be scheamtically written as :
\begin{eqnarray}
\mathcal{L} = \int d^4{\theta}\,e^{-K/3}(\tilde{K}_{T}T^*T+\tilde{K}%
_{T^c}T^{c*}T^c) + \int d^2{\theta}\,\mu_{T} {T^c}{T} +...
\end{eqnarray}
Here, $T$ and $T^c$ collectively stand for the heavy fields with masses $%
\mu_T \geq M_{\mathrm{unif}}$. In particular, they could stand for heavy $M$%
-theory states as well as heavy states in the low energy GUT multiplet (the
Higgs triplets in the $SU(5)$ case with a doublet-triplet mechanism for
example). As explained in section \ref{complete}, the heavy KK modes of the
GUT multiplet give rise to no corrections to gaugino masses. So, those modes
are not included in $T,T^c$. In the conformal supergravity formalism \cite%
{Randall:1998uk}, one introduces a conformal compensator field $C$ and
inserts it into the Lagrangian to make it conformally invariant and does
calculations. At the end, one ``gauge fixes'' the conformal compensator
field $\langle C \rangle = e^{\langle K \rangle /6}$ and $\langle
F^C/C\rangle=m_{3/2}^*-\frac{1}{3}F^m\partial_m K$ to obtain standard $%
\mathcal{N}=1,D=4$ SUGRA\footnote{%
Our definition of the moduli $F$ terms and the gauge kinetic function $f_a^0$
are slightly different from that in \cite{Choi:2007ka}.}. Here $K$ is the K%
\"{a}hler potential for the moduli and hidden sector meson field ($K =
-3\ln(4\pi^{1/3}V_X)+\bar{\phi}\phi$) in the 4d Einstein frame and $\tilde{K}%
_{T}$ is the K\"{a}hler metric for the matter field $T$. So, one writes:
\begin{eqnarray}
\mathcal{L} = \int d^4{\theta}\,CC^*\,e^{-K/3}(\tilde{K}_{T}T^*T+\tilde{K}%
_{T^c}T^{c*}T^c) + \int d^2{\theta}\,C^3\,\mu_{T} {T^c}{T} +...
\end{eqnarray}
After canonically normalizing the K\"{a}hler potential, one gets:
\begin{eqnarray}
\mathcal{L} &=& \int d^4{\theta}\,(\hat{T}^*\hat{T}+\hat{T^{c*}}\hat{T^c}) +
\int d^2{\theta}\,\frac{\mu_{T}}{\sqrt{e^{-2K/3}\tilde{K}_{T}\tilde{K}_{T}^c}%
} {\hat{T^c}}\hat{{T}} +... \\
\mathrm{where}\; \hat{T} &\equiv& (\sqrt{\tilde{K}_{T}e^{-K/3}})\,C\,T
\notag
\end{eqnarray}
So, the normalized mass of $T$ can be written as $M_{T}=\frac{\mu_{T}}{\sqrt{%
e^{-2K/3}\tilde{K}_{T}\tilde{K}_{T}^c}}$. The threshold corrections to the
gauge coupling can be written as:
\begin{eqnarray}
\Delta\,f_a^{T,T^c} = -\frac{1}{16\pi^2}\,C_a^{T}\,\ln{(\frac{M_{T}M_{T^*}}{%
M_{\mathrm{unif}}^2})}
\end{eqnarray}
Now, using (\ref{threshold}), one can write:
\begin{eqnarray}
M_{a}(M_{\mathrm{unif}}^+) - M_a(M_{\mathrm{unif}}^-) &=& g_a^2(M_{\mathrm{%
unif}})\,F^I\partial_I \,(\Delta\,f_a^{T,T^c}) \\
\implies M_a(M_{\mathrm{unif}}^-) &=& M_a(M_{\mathrm{unif}}^+) + \frac{%
g_a^2(M_{\mathrm{unif}})}{16\pi^2}\,\sum_{T,T^c} C_a^{T}\left(\frac{F^C}{C}%
+F^m \partial_m\ln(e^{-2 K/3}{\tilde K}_{T} {\tilde K}_{T^c})\right)  \notag
\\
\mathrm{where}\; M_a(M_{\mathrm{unif}}^+) &=& g_a^2(M_{\mathrm{unif}})[\frac{%
1}{8\pi}F^m\partial_m f_a^0+\frac{1}{8\pi^2}\sum_{MSSM,T,T^c}
C^i_aF^m\partial_m\ln{(e^{-K/3}\tilde{K}_i)}-  \notag \\
& & \frac{1}{16\pi^2}(3C_a-\sum_{MSSM,T,T^c}C_a^i)\frac{F^C}{C}]  \notag
\end{eqnarray}
So, one finally gets for $M_a(M_{\mathrm{unif}}^-)$:
\begin{eqnarray}  \label{gauginoT}
M_a(M_{\mathrm{unif}}^-) &=& g_a^2(M_{\mathrm{unif}})\,[\frac{1}{8\pi}%
F^m\partial_m f_a^0+\frac{1}{8\pi^2}\sum_{i=1}^{MSSM} C^i_aF^m\partial_m\ln{%
(e^{-K/3}\tilde{K}_i)}-\frac{1}{16\pi^2}(3C_a-\sum_{i=1}^{MSSM}C_a^i)\frac{%
F^C}{C}]  \notag \\
&=& \frac{g_a^2(\mu)}{8\pi}\,\left(F^m\partial_m\,f_a^0\right)-\frac{%
g_a^2(\mu)}{16\pi^2}\Big(-(3C_a-\sum_{i}C_a^{i})
e^{K/2}W^{*}+(C_a-\sum_{i}C_a^{i})F^mK_m  \notag \\
& &+2\sum_{i=1}^{MSSM}(C_a^{i}F^m\partial_m\ln(\tilde{K}_i))\Big)  \notag \\
&=& M_a^{tree}(M_{\mathrm{unif}}^-)+M_a^{anom}(M_{\mathrm{unif}}^-)
\end{eqnarray}
where $i$ runs only over the MSSM particles. Just above the unification
scale, the beta function coefficients corresponded to that of the MSSM and
the heavy fields $T,T^c$. From (\ref{gauginoT}), we see that the threshold
correction to the gaugino masses because of the heavy fields exactly cancels
the heavy field contribution to $M_a(M_{\mathrm{unif}}^+)$! \emph{This
implies that below the unification scale, we can forget about the heavy
fields $T,T^c$ and just take effects of the MSSM particles into account.}

\end{document}